\documentclass[3p]{elsarticle}

\usepackage{lineno,hyperref}

\journal{Journal of Computer Methods in Applied Mechanics and Engineering}

\usepackage[linesnumbered,ruled]{algorithm2e}
\usepackage{amsmath}
\usepackage{amssymb}
\usepackage{bm}
\usepackage{cases}
\usepackage{graphicx}
\usepackage{mathtools}
\usepackage{multirow}
\usepackage{subcaption}
\usepackage{xcolor}

\definecolor{corrections}{RGB}{0, 110, 240} 
\newcommand{\reviewed}{\textcolor{black}}	

\DeclarePairedDelimiter{\ceil}{\lceil}{\rceil}

\begin{document}

\nolinenumbers

\begin{frontmatter}

\title{Patch-wise Quadrature of Trimmed Surfaces in Isogeometric Analysis}

\author[1]{Michael Loibl\corref{mycorrespondingauthor}}
\ead{michael.loibl@unibw.de}
\author[2]{Leonardo Leonetti}
\author[3]{Alessandro Reali}
\author[1]{Josef Kiendl}
\address[1]{Institute of Engineering Mechanics and Structural Analysis, Universität der Bundeswehr München, Werner-Heisenberg-Weg 39, 85577 Neubiberg, Germany}
\address[2]{Dipartimento di Ingegneria Informatica, Modellistica, Elettronica e Sistemistica, Università della Calabria, Italy}
\address[3]{Department of Civil Engineering and Architecture, University of Pavia, Italy}

\cortext[mycorrespondingauthor]{Corresponding author}

\begin{abstract}
This work presents an efficient quadrature rule for shell analysis \reviewed{fully integrated in CAD} by means of Isogeometric Analysis (IGA). General CAD-models may consist of trimmed parts such as holes, intersections, cut-offs etc. Therefore, IGA should be able to deal with these models in order to fulfil its promise of closing the gap between design and analysis. Trimming operations violate the tensor-product structure of the used Non-Uniform Rational B-spline (NURBS) basis functions and of typical quadrature rules. Existing efficient patch-wise quadrature rules consider actual knot vectors and are determined in 1D. They are extended to further dimensions by means of a tensor-product. Therefore, they are not directly applicable to trimmed structures. The herein proposed method extends patch-wise quadrature rules to trimmed surfaces. Thereby, the number of quadrature points can be significantly reduced. Geometrically linear and non-linear benchmarks of plane, plate and shell structures are investigated. The results are compared to a standard trimming procedure and a good performance is observed.
\end{abstract}

\begin{keyword}
patch-wise quadrature\sep generalized Gaussian quadrature\sep trimming\sep shells\sep Isogeometric Analysis
\end{keyword}

\end{frontmatter}


\section{Introduction}
When Isogeometric Analysis (IGA) was first presented in 2005, it promised to close the gap between design and analysis \cite{Hughes2005}. The classical Finite Element Method (FEM) adopts different functions in order to describe geometries than the ones used in Computer Aided Design (CAD). This hinders a direct communication between design process and accompanying analysis. The key idea of IGA is therefore to directly employ the functions used in CAD as basis functions for the analysis. The most established functions in CAD as well as in IGA are Non-Uniform Rational B-Splines (NURBS) which enable a broad flexibility in modelling, e.g. they are able to exactly represent conic sections, which include circles and ellipses \cite{Hughes2005}. \bigskip

The reduction of computational costs is of general interest in the context of numerical simulations. A good overview over different methods to speed-up computations in IGA is given by \cite{Pan2020}. Four approaches can be distinguished: 
\begin{enumerate}
	\item Collocation methods
	\item Quadrature techniques employing sum factorization
	\item Quadrature-free approaches
	\item Reduced or specialized quadrature rules
\end{enumerate}
Collocation methods do not solve the weak form of a PDE, but enforce its strong form at a set of locations (called collocation points) \cite{Auricchio2010}. Sum factorization reduces the computational cost by splitting up the tensor-product of the basis functions and, thereby, computing the integral one direction after the other \cite{Antolin2015}. Quadrature-free approaches completely abandon numerical integration by applying interpolation of the integrands and precomputed look-up tables \cite{Pan2020}. \bigskip

In this work, the focus is on specialized quadrature rules. A standard ``$p$+1'' Gauss rule is typically used in IGA for the integration of stiffness matrices. NURBS normally inherit a high smoothness and span over multiple elements (knot spans). Therefore, a standard Gauss rule uses more quadrature points than necessary because it does not consider the higher inter-element smoothness. In the last decade, many specialized quadrature rules were developed providing an optimal number of quadrature points by considering the actual parametrization of the patch. These rules are often called patch-wise quadrature rules \cite{Teschemacher2018,Leonetti2018} or generalized Gaussian quadratures \cite{Calabro2019}. The first patch-wise quadrature rule was presented by \cite{Hughes2010} and is called ``half-point rule'' referring to the fact that the number of quadrature points is roughly equal to half the number of basis functions of the target space under consideration. The rule is independent of the polynomial order. However, it does not consider the effects of the boundaries and of non-uniform regularities. Auricchio et al. \cite{Auricchio2012} proposed a robust, almost optimal rule considering boundary effects which is however restricted to uniform knot vectors and regularities. Adam et al. \cite{Adam2015} presented an algorithm which finds local optimal quadrature points and, thus, is able to develop a patch-wise rule in a very stable element-wise manner. Johannessen \cite{Johannessen2017} focused on developing a rule for completely arbitrary knot spans and polynomial degrees. His main contributions were the determining of good initial values and the development of a stable recursive algorithm. A similar contribution - however considering uniform regularities - was published by \cite{Hiemstra2017} where quadrature points for the most common spline space were explicitly documented. Calabrò et al. \cite{Calabro2017} proposed the so called Weighted Quadrature. Hereby, the idea is that the test functions are incorporated in the weights. This rises the possibility that the integrals are computed for one basis function after another instead of an element-wise computation. This approach used in combination with other methods \reviewed{(e.g. fast formation and assembly)} showed significant gains with respect to the computational efficiency \cite{Sangalli2018,Hiemstra2019,Giannelli2021}. However, it makes typical FEM-like implementations for IGA unusable because it abandons the element-wise assembly. Furthermore, this quadrature rule leads to not perfectly symmetric matrices (e.g. stiffness or mass matrices). Thus, it imposes the necessity of non-symmetric solvers. Selective and reduced patch-wise quadrature rules were presented by \cite{Adam2015} in order to prevent volumetric locking. Avoiding membrane locking in the context of thin shells by means of reduced patch-wise quadrature was studied by \cite{Leonetti2019}. Hokkanen et al. \cite{Hokkanen2020} pointed out certain problems of reduced patch-wise quadrature with still present locking and with hourglassing in the context of more complex (real-world) shells. Hokkanen \cite{Hokkanen2019} compared different patch-wise quadrature rules and showed the severe impact of these methods on the computational time. This holds in particular for non-linear analyses where the computations at the single quadrature points are very often repeated. Calabrò et al. \cite{Calabro2019} provide a good overview with further details about the developments of patch-wise quadrature rules in IGA. \bigskip

More complex CAD-models are often created by means of trimming operations, which facilitate a large variety of geometries including for example structures with arbitrary intersections and voids. Therefore, it is important to be able to deal with trimmed structures in IGA. In the present work, the focus lies on surface structures. Most trimmed CAD surface models can be described by its Boundary Representation (B-Rep). Other geometric representations such as Constructive Solid Geometries (CSG) are easily convertible to B-Rep models \cite{Breitenberger2016}. This gave rise to the concept of the Isogeometric B-Rep Analysis (IBRA), which introduced a detailed framework for simulations of trimmed models \cite{Teschemacher2018,Breitenberger2015}. A good review about the different trimming techniques in the context of IGA is provided by \cite{Marussig2018}. Trimming procedures in IGA can be distinguished in local and global approaches. The global strategies fix the patch by re-parametrization or reconstruction. Thereby, the initial CAD-geometry is not maintained. This contradicts the philosophy of integrating design and analysis by using the same model. Following \cite{Marussig2018}, three different local strategies can be distinguished: Local reconstruction, approximated trimming curve and exact trimming curve. The first strategy locally reconstructs trimmed elements with single patches \cite{Schmidt2012}. The second group can be further distinguished in tailored integration rules and adaptive subdivision schemes. Tailored quadrature rules determine the quadrature points by solving a moment-fitting equation where the reference result stems from a line integral \cite{Nagy2015}. In particular, an iterative point elimination procedure was developed in \cite{Nagy2015} in order to solve this highly non-linear problem. This was applied to trimmed solids in \cite{Messmer2022}. Adaptive subdivision, such as the Finite Cell Method, divides trimmed elements in a tree-like manner until a certain resolution \cite{Rank2012}. The third group maps quadrature points (mostly Gauss points) from an integration space to the parameter space of a trimmed element considering the exact trimming curve \cite{Breitenberger2016,Kim2009,Kim2010,Guo2018}. \bigskip

It would be favourable to apply the discussed patch-wise quadrature rules to trimmed structures to utilize the gains of efficiency also in case of general CAD-models. The key problem is that these patch-wise quadrature rules are constructed in 1D and that they are extended to further dimensions by a tensor-product. However, the tensor-product structure is violated by trimming operations. In the present work, a method is proposed which enables the use of patch-wise quadrature rules for trimmed surfaces. The presented approach is computationally more efficient compared to standard Gaussian quadrature by reducing the overall number of quadrature points. It is highlighted that the proposed method can be combined with any patch-wise quadrature rule which is based on element-wise assembly and any local trimming technique. In this work, the patch-wise quadrature rule of \cite{Johannessen2017} is used because of its applicability to arbitrary knot vectors and polynomial degrees as well as its numerical stability. This rule was successfully applied in different structural simulations (e.g. \cite{Leonetti2019,Leonetti2018a}) Furthermore, a trimming technique which considers the exact trimming curve is used following the procedure by \cite{Kim2009} for the distinction of the trimming cases and applying the blending function method described in \cite{Guo2018} for determining the quadrature points of the trimmed elements. Marussig \cite{Marussig2022} presented an extension of the Weighted Quadrature to trimmed plane surfaces. Meßmer et al. \cite{Messmer2022} proposed the application of patch-wise quadrature rules to trimmed solids where they grouped the knot spans to untrimmed macro-elements for which they used more efficient patch-wise rules. In contrast to these two works, the focus is here on general trimmed surfaces (plates and shells) using element-wise assembly. Instead of using macro-elements, a transition zone is introduced which enables the use of a patch-wise rule determined for the complete patch and which provides higher flexibility of the method with respect to arbitrary trimming curves. \bigskip

The paper is organized as follows. In Section \ref{sec:nurbs}, the concept of B-splines and Non Uniform Rational B-splines (NURBS) is presented and the subsequently used terminology is introduced. Section \ref{sec:numerical_quadrature} describes numerical quadrature in the context of IGA. Special focus is given to patch-wise quadrature rules and the integration of trimmed elements in Subsections \ref{sec:patch_wise_integration} and \ref{sec:integration_trimmed}, respectively. The proposed extension of the patch-wise rules to trimmed surfaces is outlined in Section \ref{sec:patch_wise_integration_trimmed_surface}. In the subsequent Section \ref{sec:numerical_results}, several numerical benchmarks are computed. Thereby, the efficiency gains with respect to the reduction of quadrature points of the proposed method are demonstrated and its applicability to a range of different trimming problems is shown. The final Section \ref{sec:conclusion} concludes the paper and provides an outlook to further research.

\section{NURBS Basis Functions} \label{sec:nurbs}
This section presents a short introduction to NURBS, which are most often used in IGA as basis functions. A more detailed description as well as algorithms for the most important geometric operations in the context of B-splines and NURBS can be found in \cite{Piegl1995}. NURBS are an extension of B-splines. A B-spline curve is defined as
\begin{equation}
	\mathbf{C}(\xi) = \sum_{i=1}^{n}N_{i,p}(\xi)\mathbf{P}_i
\end{equation}
where $\xi$ is the curve parameter, $P_i$ the $i$\textsuperscript{th} control point, $n$ the number of control points and $N_{i,p}$ the B-spline basis function of polynomial degree $p$ corresponding to control point $i$. B-splines are described by piecewise polynomial segments of degree $p$. Thereby, adjacent segments are joined at the knots $\xi_i$. The knots form a non-decreasing sequence of parametric coordinates. They are collected in a knot vector $\Xi$:
\begin{equation}
	\Xi = \{\xi_1,\xi_2,...,\xi_m\}, \quad \xi_{i} \leq \xi_{i+1}
\end{equation}
where the size of the knot vector is $m$. The number of control points can be computed as
\begin{equation} \label{equ:n_common}
	n = m - p - 1
\end{equation}

The continuity of a B-spline at each knot is equal to $C^{p-k}$ where $k$ is the knot multiplicity. Open knot vectors are normally used, which means that the first and the last knot is repeated ``$p$+1''-times. This yields as a practical consequence that the first and the last control point are interpolated and that the curve tangents at these points have the same direction as the control polygon. Uniform continuity is defined as an open knot vector where all inner knots have the same multiplicity $k$. The continuity at inner knots is called regularity $r=p-k$. Knots with maximum regularity ($k=1$) have a continuity of $C^{p-1}$. Eq.~\eqref{equ:n_common} may be reformulated under the assumption of uniform continuity as
\begin{equation}\label{equ:n}
	n = (p+1)n_{ele} - (r+1)(n_{ele}-1)
\end{equation}
where $n_{ele}$ denotes the number of non-zero knot spans, which is noted as element in IGA. B-splines satisfy all properties of a basis and span a solution space:
\begin{equation}
	\mathbb{S}^p_r = \text{span}\{{N_{i,p,\Xi}}\}
\end{equation}

B-splines can be extended to 2D in a tensor-product manner:
\begin{equation}
	N_{i,j}^{p,q} = N_{i,p}(\xi)M_{j,q}(\eta)
\end{equation}
where $M_{j,q}$ is a B-spline defined in a second parametric direction $\eta$ with polynomial degree $q$. NURBS are derived from B-splines as:
\begin{equation}\label{equ:NURBS}
	R_{i,p}(\xi) = \dfrac{N_{i,p}(\xi)w_i}{\sum_{i=1}^n N_{i,p}(\xi)w_i}
\end{equation}
where $w_i$ is an additional weigth corresponding to the $i$\textsuperscript{th} control point. NURBS are extended to 2D by a weighted tensor-product:
\begin{equation}
	R_{i,j}^{p,q} = \dfrac{N_{i,p}(\xi)M_{j,q}(\eta)w_{i,j}}{\sum_{i=1}^{n}\sum_{j=1}^{m}N_{i,p}(\xi)M_{j,q}(\eta)w_{i,j}}
\end{equation}
If the control points in both directions are collected and sequentially numbered, \reviewed{the definition of a NURBS surface reads:}
\begin{equation}
	\mathbf{S}(\bm{\xi}) = \sum_{i=1}^n R_{i}(\bm{\xi}) \mathbf{P}_{i}
\end{equation}
where $\bm{\xi} = (\xi,\eta)$ are the parametric coordinates\reviewed{, $\mathbf{P}_i$ is the $i$\textsuperscript{th} control point} and $n$ is now the total number of control points.

\section{Numerical Quadrature in Isogeometric Analysis} \label{sec:numerical_quadrature}
The solution of the weak form of a mechanical problem in IGA requires the computation of integrals. A simple FEM-inspired approach is to use Gaussian quadrature on each element \cite{Hughes2005}. This implies the assumption that the functions which are integrated can be treated as polynomials. Calabr\`o et al. state that the optimal order of convergence is maintained \reviewed{for NURBS basis functions} when applying Gaussian quadrature (see \cite{Calabro2019} and references therein). The order of the Gauss rule depends on the polynomial degree of the integrand. In the context of this work, we are focusing on static, structural computations. Therefore, the quadrature rule should be able to compute stiffness matrices. A full Gaussian quadrature means that ``$p$+1'' quadrature points per element and direction are used to integrate stiffness terms, where $p$ is the polynomial order of the basis functions.

\subsection{Patch-wise Quadrature} \label{sec:patch_wise_integration}
In contrast to the standard Gauss rule, patch-wise quadrature considers the higher regularity of the NURBS basis functions (they are defined over multiple elements). Therefore, these quadrature rules are more efficient in the sense that they need less quadrature points. These rules depend on the actual integrand which is considered in the weak formulation. In this work, the static analysis of plane (plane stress and plane strain) and Kirchhoff-Love shell models is considered, which corresponds to the solution of partial differential equations of $2^{\text{nd}}$ and $4^{\text{th}}$ order, respectively. A detailed description of the used shell element can be found in \cite{Kiendl2011}. The basis functions are tensor-product functions and the involved integrals are sequential integrations of the parametric directions. Therefore, it is possible to derive quadrature rules in 1D and to extend them to more dimensions in a tensor-product manner. The weak form of a $2^{\text{nd}}$ order differential equation involves terms such as
\begin{equation}\label{equ:integral_2D} 
	\int_{\Omega} N_i'(\xi) N_j'(\xi) d\Omega
\end{equation}
where $\Omega$ is the integration domain, and $N_i'$ and $N_j'$ are the first derivatives of the B-spline basis functions corresponding to the $i$\textsuperscript{th} and $j$\textsuperscript{th} control point, respectively. The weak form of a $4^{\text{th}}$ order differential equation involves terms such as
\begin{equation}\label{equ:integral_KL} 
	\int_{\Omega} N_i''(\xi) N_j''(\xi) d\Omega
\end{equation}
where $N_i''$ and $N_j''$ are the second derivatives of the B-spline basis functions. The plane element requires first order derivatives of the shape functions as indicated in Eq.~\eqref{equ:integral_2D} and the shell element requires second order derivatives of the shape functions as indicated in Eq.~\eqref{equ:integral_KL}. Even though this paper focuses on these two specific element types in the context of static analysis, the general ideas are applicable to other problems described by $2^{\text{nd}}$ and $4^{\text{th}}$ order partial differential equations. Rational terms (e.g. arising from the determinant of the Jacobian or the NURBS weights) are normally omitted in the derivation of quadrature rules in the context of IGA (see \cite{Hughes2010} for a good reasoning). \bigskip

The function space considering the actual integrand is called target space, i.e. considering Eqs.~\eqref{equ:integral_2D} and~\eqref{equ:integral_KL}. The target spaces of the plane and the Kirchhoff-Love element are listed in Table \ref{tab:target_space}. The continuity of the target spaces arises from the terms of highest order in the integrands given by a multiplication of two basis functions of degree $p$ (therefore, the continuity is $2p$). The regularity of the target spaces is defined by the highest derivative involved as indicated by Eqs.~\eqref{equ:integral_2D} and~\eqref{equ:integral_KL}. The function space which is considered in order to develop a quadrature rule is called an approximation space and does not necessarily coincide with the target space \cite{Adam2015}. A quadrature rule which takes the actual target space as approximation space is called full patch-wise quadrature rule. The standard Gauss rule for example corresponds to an approximation space $\mathbb{S}^{2p}_{-1}$. The actual target spaces given in Table \ref{tab:target_space} are subspaces of this space ($\mathbb{S}^{2p}_{r-1} \subset \mathbb{S}^{2p}_{r-2} \subset \mathbb{S}^{2p}_{-1}$ with $r>2$). A function space is a subspace if they have a higher regularity and/or a lower degree. \bigskip

\begin{table} 
	\caption{\label{tab:target_space}Target spaces for plane element and Kirchhoff-Love shell element for the solution space $\mathbb{S}_r^p$}
	\begin{center}
		\begin{tabular}{cc}
			Element & Target space \\ \hline &\\[-2ex]
			Plane & $\mathbb{S}_{r-1}^{2p}$ \\
			Kirchhoff-Love shell & $\mathbb{S}_{r-2}^{2p}$
		\end{tabular}
	\end{center}
\end{table}

\begin{figure} [!bth]
	\centering
	\includegraphics{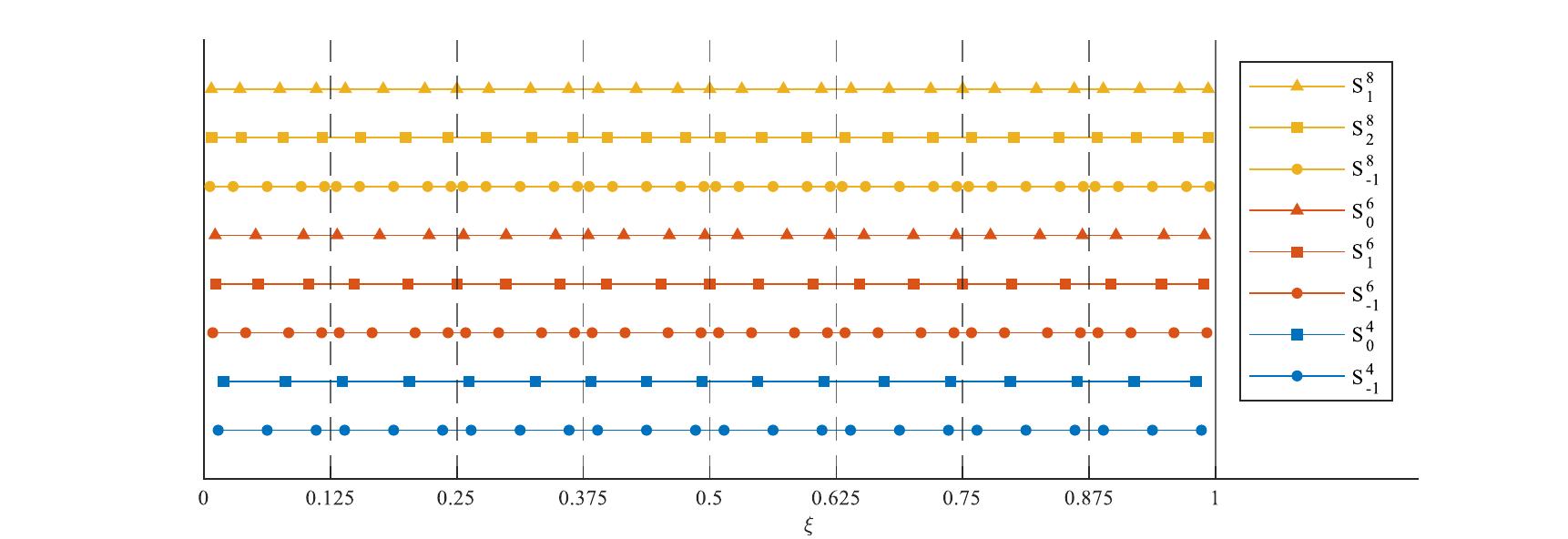}
	\caption{Quadrature points in 1D due to Gaussian (circles) and full patch-wise quadrature rules with respect to different polynomial degrees shown for a knot vector $\Xi = \{0,...,0,0.125,0.25,0.375,0.5,0.625,0.75,0.875,1,...,1\}$ of the solution space. The target spaces of a plane (squares) and a Kirchhoff-Love shell (triangles) element are considered.}
	\label{fig:comparison_Gauss_pw}
\end{figure}

Patch-wise quadrature deals with the actual target spaces as listed in Table \ref{tab:target_space}. A numerical quadrature rule is considered exact if it fulfils
\begin{equation}\label{equ:integration}
	\sum_i^{n_{quad}} w_i f(x_i) := \int_{\mathbb{R}} f(x) dx
\end{equation}
where $f$ is the integrand, $x$ the variable of the function $f$, $n_{quad}$ the number of quadrature points, $w_i$ the integration weight of the $i$\textsuperscript{th} quadrature point and $x_i$ the position of the $i$\textsuperscript{th} quadrature point. The generic Eq.~\eqref{equ:integration} contains only a single variable $x$ because patch-wise quadrature rules are normally derived with respect to an uni-variate integral (1D). For a set of functions this equation can be rewritten in matrix notation as
\begin{equation}\label{equ:system_of_equations}
	\left[\begin{array}{cccc}
		f_1(x_1) &f_1(x_2) &\hdots &f_1(x_{n_{quad}}) \\
		f_2(x_1) &f_2(x_2) &\hdots &f_2(x_{n_{quad}}) \\
		\vdots & &\ddots &\vdots \\
		f_n(x_1) &f_n(x_2) &\hdots &f_n(x_{n_{quad}}) \\
	\end{array}\right] \left[\begin{array}{c}
		w_1 \\
		w_2 \\
		\vdots \\
		w_{n_{quad}}
	\end{array}\right] = 	
	\left(\begin{array}{c}
	\int f_1(x)dx \\
	\int f_2(x)dx \\
	\vdots \\
	\int f_n(x)dx
\end{array}\right)
\end{equation}
where $n$ denotes the number of functions in the considered function space and $n_{quad}$ denotes the number of quadrature points. In the context of full patch-wise quadrature, the parameter $n$ refers to the number of uni-variate NURBS basis functions used for the discretization of the target space. The system of equations~\eqref{equ:system_of_equations} has $2n_{quad}$ unknowns (weights and positions of the quadrature points) and $n$ equations. Therefore, the minimum (optimal) number of quadrature points in order to exactly integrate the function space is
\begin{equation} \label{equ:n_quad}
	n_{quad} = \ceil[\Big]{\dfrac{n}{2}}
\end{equation}
where $\lceil \cdot \rceil$ is the nearest integer larger or equal to $x$. In this work, the patch-wise quadrature rule by \cite{Johannessen2017} is used to determine the positions and weights of the quadrature points in Eq.~\eqref{equ:system_of_equations}. However, the proposed method is not restricted to this explicit quadrature rule. In order to derive an estimate for the number of quadrature points with the patch-wise quadrature rule, Eq.~\eqref{equ:n_quad} can be applied to the target spaces from Table \ref{tab:target_space} making use of Eq.~\eqref{equ:n} and assuming maximum regularity. Considering plane and Kirchhoff-Love (KL) shell elements, it yields
\begin{equation}\label{equ:n_quad_2D}
	n_{quad,plane} = \dfrac{dim(\mathbb{S}_{r-1}^{2p})}{2} = \dfrac{(p+2)n_{ele}+p-1}{2} = \mathcal{O}(\dfrac{p+2}{2}n_{ele})
\end{equation}
\begin{equation}\label{equ:n_quad_KL}
	n_{quad,KL} = \dfrac{dim(\mathbb{S}_{r-2}^{2p})}{2} = \dfrac{(p+3)n_{ele}+p-2}{2} = \mathcal{O}(\dfrac{p+3}{2}n_{ele})
\end{equation}
It should be noted that neither the patch-wise quadrature rule by \cite{Johannessen2017} nor the present work are restricted to certain regularities, but the assumption of maximum regularity is only made to derive the estimates in Eq.~\eqref{equ:n_quad_2D} and~\eqref{equ:n_quad_KL}. In particular, the possibility to deal with arbitrary regularities is important if the direct use of CAD-models is desired \cite{Dornisch2021}. It is observed that the number of quadrature points per element levels asymptotically with \reviewed{element refinement}. Table \ref{tab:reduction_n_quad} shows these asymptotic numbers of quadrature points per element in 1D. The full patch-wise rules reduce to Gauss rules in case of $p=1$ and $p=2$ for the plane and the Kirchhoff-Love element, respectively. The Kirchhoff-Love element cannot be used with $p=1$ because it requires at least $C^1$ continuity. The reduction of quadrature points compared to a Gauss rule levels asymptotically to 0.5 with increasing $p$. The Fig. \ref{fig:comparison_Gauss_pw} shows the quadrature points exemplary for an open knot vector with eight elements and maximum regularity. \bigskip
\begin{table} 
	\caption{\label{tab:reduction_n_quad}Asymptotic limit of the number of quadrature points per element per direction for different polynomial degrees $p$ of the solution space with respect to different quadrature rules and element types considering maximum regularity. The numbers in brackets give the reduction of points comparing the patch-wise and the Gaussian quadrature rules.}
	\begin{center}
		\begin{tabular}{cccc}
			\multirow{2}{*}{p} & \multirow{2}{*}{Gauss rule} & \multicolumn{2}{c}{Full patch-wise rules} \\
			&& Plane element &  KL shell element \\ \hline &&&\\[-2ex]
			1 & 2 & not applicable & not applicable \\
			2 & 3 & 2 (0.67) & not applicable \\
			3 & 4 & 2.5 (0.63) & 3 (0.75) \\
			4 & 5 & 3 (0.60) & 3.5 (0.70) \\
			5 & 6 & 3.5 (0.58) & 4 (0.67)
		\end{tabular}
	\end{center}
\end{table}

It is important to notice that a patch-wise quadrature rule partly overcomes the element-wise thinking typical for FEM. The reason is that the quadrature points within one element do not longer integrate solely over the respective knot span. In FEM, an element stiffness matrix is computed by an integration over the element. Whereas, the here presented patch-wise quadrature rule abolishes the definition of the element bounds as integration bounds. Therefore, it is no longer possible to solely compute the contribution of a single knot span to the stiffness matrix. This plays an important role when the described patch-wise rule should be extended to the case of a trimmed surface. The described behaviour can be observed by summing up the weights of the quadrature points lying within one element. It is normally expected that this yields the element area because the Gauss points are mapped from the integration space onto the element. However, this is not the case for patch-wise quadrature rules.

\subsection{Quadrature of Trimmed Elements}\label{sec:integration_trimmed}
\begin{figure} [!bth]
	\centering
	\includegraphics[width=\textwidth]{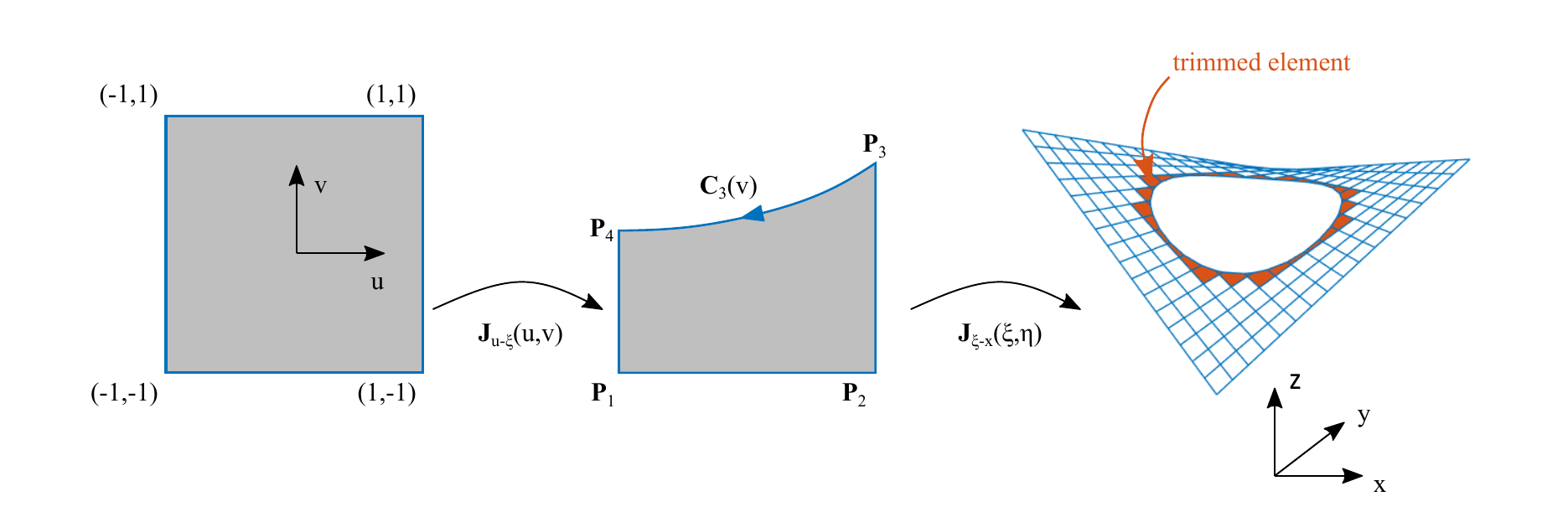}
	\caption{Mapping of a trimmed element between the integration, the parametric and the physical space (from left to right).}
	\label{fig:trimmed_element_mapping}
\end{figure}

The integration domain of an element $i$ on a trimmed surface is defined by the intersection of the element area $\Omega_i$ with the valid domain $\Omega^v$, i.e. $\mathbf{S}_i^v := \Omega_i \cap \Omega^v$. Three different types of elements can be distinguished based on the specification of their integration domains:
\begin{itemize}
	\item \textit{Active-untrimmed} if $\mathbf{S}_i^v = \Omega_i$
	\item \textit{Trimmed} if $0 < \mathbf{S}_i^v < \Omega_i$
	\item \textit{Inactive} if $\mathbf{S}_i^v = \emptyset$
\end{itemize}
An example for this distinction is found in Fig. \ref{fig:element_numbering}. In the present work, a local trimming technique which considers parametric trimming curves is used. The blending function method is used to map Gauss quadrature points onto trimmed elements \cite{Guo2018}. However, arbitrary local trimming techniques can be applied within the here proposed method of patch-wise quadrature of trimmed surfaces. The idea of the blending function method is a map from an integration space to a parametric space applying a linear blending function, whereby any of the four sides of the parametric element can be curved. Detailed formulas of the method can be found in \cite{Szabo1991}. The number of quadrature points, which are mapped, is determined by the ``$p+1$''-rule considering the highest occurring polynomial degree of the trimming curve and the surface basis functions as it is suggested by \cite{Breitenberger2016}. Fig. \ref{fig:trimmed_element_mapping} shows the mapping of a trimmed element between the integration, the parametric and the physical space. In this case, the third edge is curved. However, the method is also able to deal with three-sided elements. Trimmed elements with more edges have to be subdivided into quadrilaterals or triangles. This is also the case for other local trimming techniques \cite{Breitenberger2016,Marussig2018,Kim2009,Kim2010,Guo2018}. In general, the subdivision leads to an increased number of quadrature points in trimmed elements. For further details about the here used trimming procedure, the reader is referred to \cite{Kim2009,Kim2010}.

\section{Patch-wise Quadrature of Trimmed Surfaces} \label{sec:patch_wise_integration_trimmed_surface}
\begin{figure} [!bth]
	\centering
	\makebox[\textwidth][c]{\includegraphics[width=1.2\textwidth]{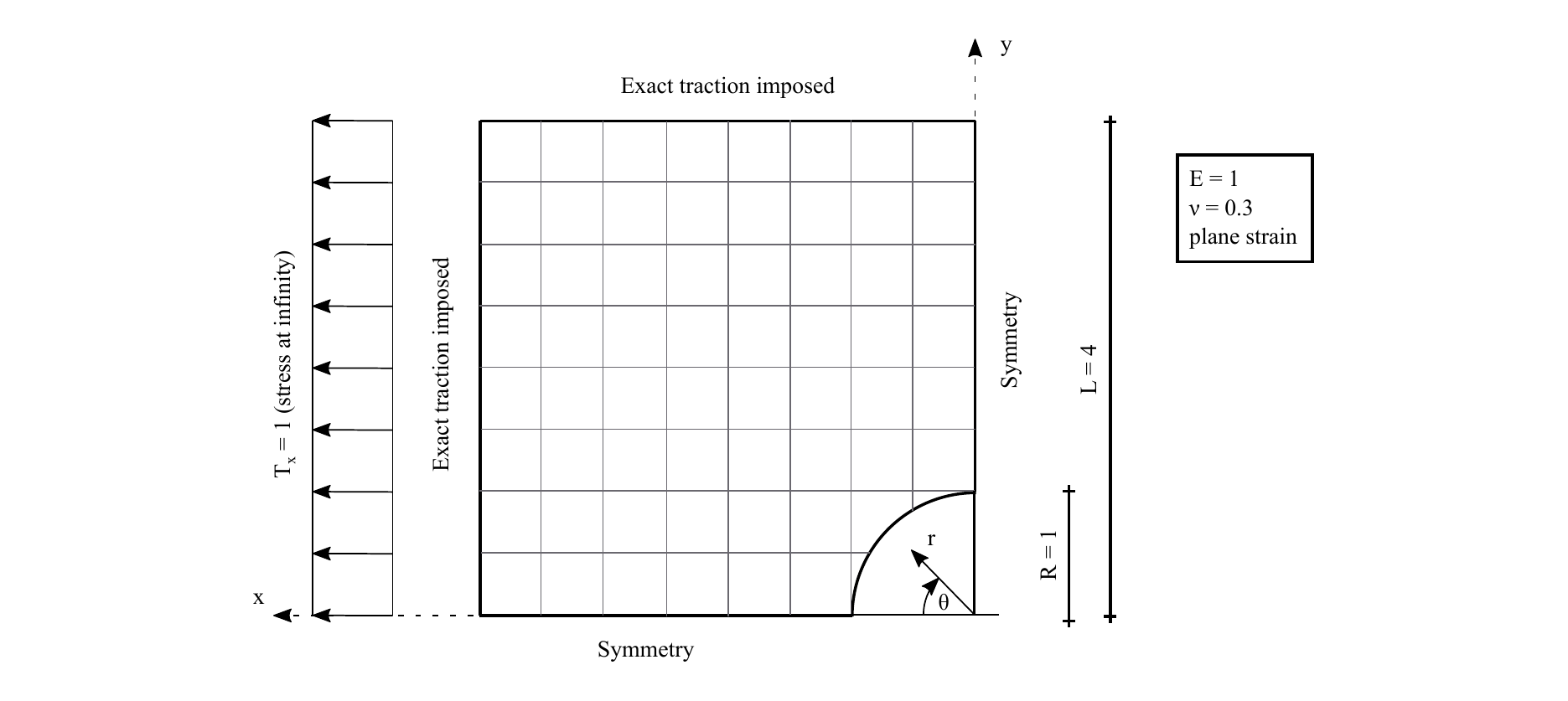}}
	\caption{Setup of the infinite plate with circular hole.}
	\label{fig:infinite_plate_with_circle}
\end{figure}

In this section, we present the extension of the patch-wise quadrature rule outlined in Section \ref{sec:patch_wise_integration} to trimmed surfaces. The basic concept of our method is explained in Section \ref{sec:concept}. Afterwards, the gain with respect to the computational efficiency, which is the main motivation for enhanced quadrature rules, is discussed in Section \ref{sec:efficiency_considerations}. The following explanations are visualized by the well-known example of an infinite plate with a circular hole \cite{Hughes2005}. The same example is later used for some numerical tests. The problem description is given in Fig. \ref{fig:infinite_plate_with_circle}. All explanatory figures are based on a $8\times8$ mesh and polynomial degrees $p=q=2$.

\subsection{Concept} \label{sec:concept}
\begin{figure} [!bth]	
	\centering
	\begin{subfigure}{0.47\textwidth}
		\centering
		\includegraphics{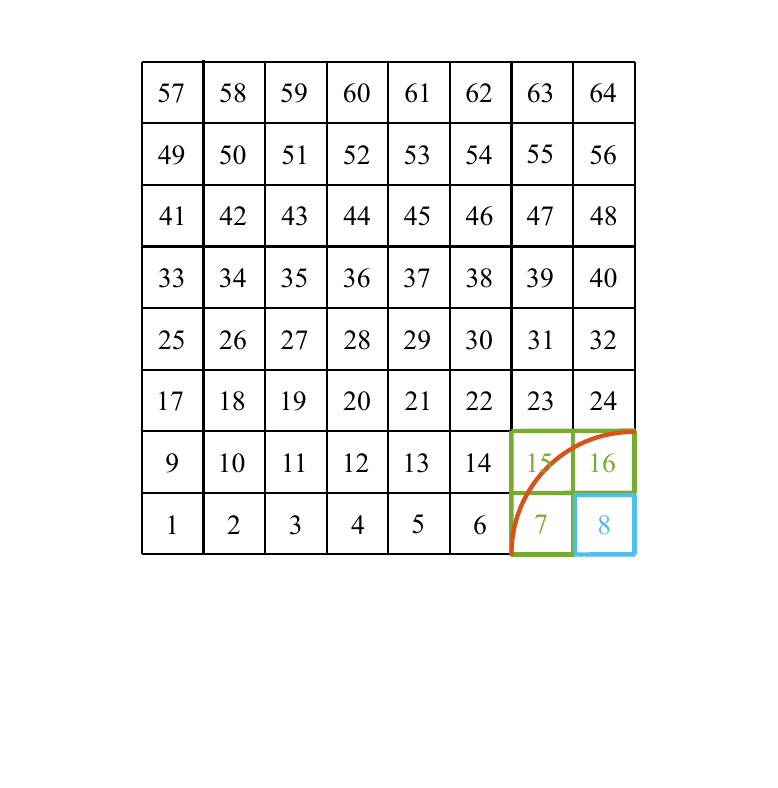}
		\caption{Element numbering and different element types in trimming distinguished by color. Black = \textit{active-untrimmed}, green = \textit{trimmed}, light blue = \textit{inactive}.}
		\label{fig:element_numbering}
	\end{subfigure}\hfill
	\begin{subfigure}{0.47\textwidth}
		\centering	
		\includegraphics{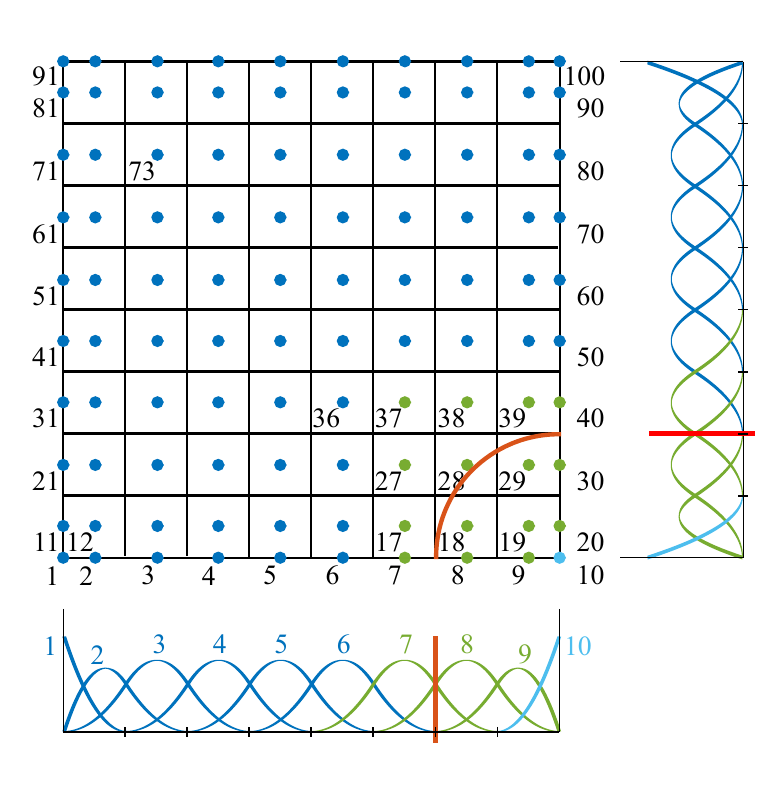}
		\caption{Basis function numbering and different basis function types in trimming distinguished by color. Dark blue = \textit{active-untrimmed}, green = \textit{trimmed}, light blue = \textit{inactive}.}
		\label{fig:basis_function_distinction}
	\end{subfigure}\hfill
	\begin{subfigure}{0.47\textwidth}
		\centering
		\includegraphics{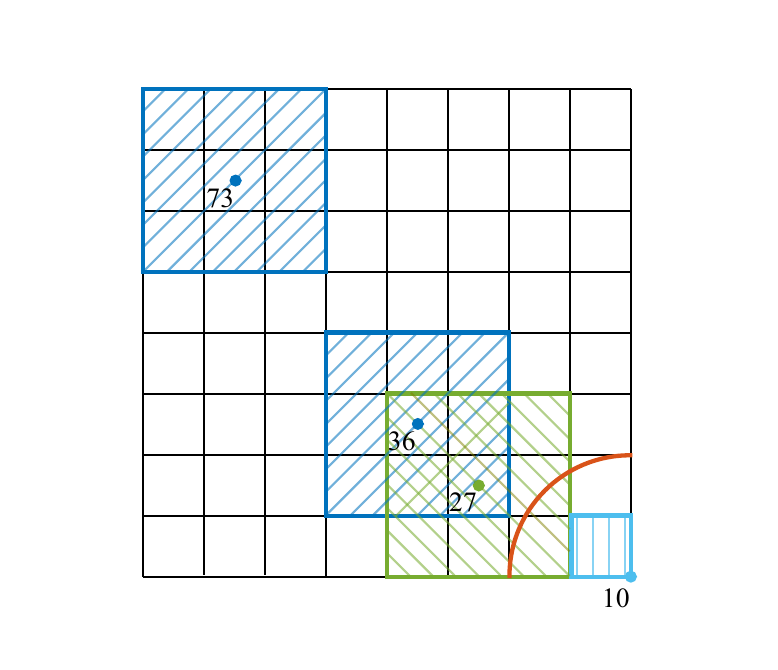}
		\caption{Support of selected basis functions where the basis functions are distinguished by colors and hatching. Dark blue with 45°-hatching = \textit{active-untrimmed}, green with 135°-hatching = \textit{trimmed}, light blue with vertical hatching = \textit{inactive}.}
		\label{fig:basis_function_support}
	\end{subfigure}\hfill
	\begin{subfigure}{0.47\textwidth}
		\centering
		\includegraphics{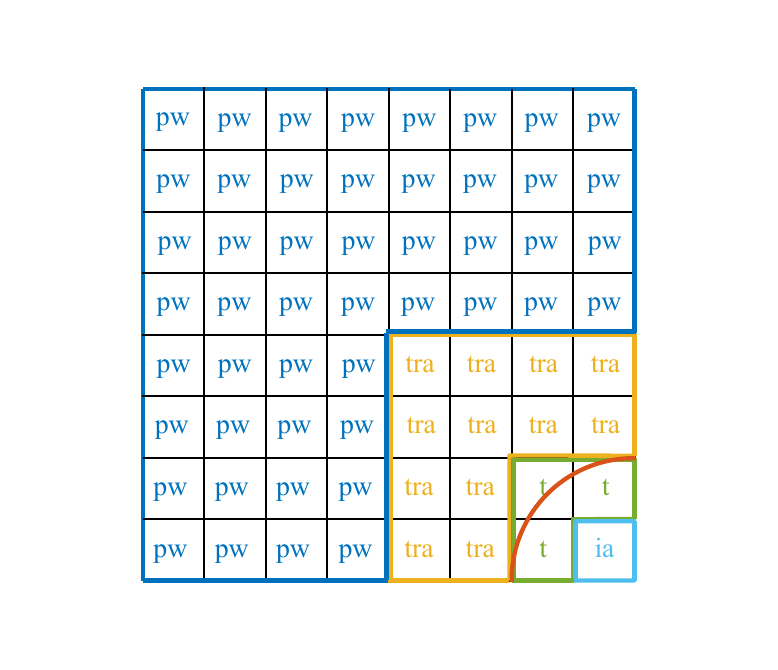}
		\caption{Element distinction in the proposed method indicated by colors and abbreviations. \textit{pw} in dark blue = \textit{patch-wise}, \textit{tra} in gold = \textit{transition}, \textit{t} in green = \textit{trimmed}, \textit{ia} in light blue = \textit{inactive}.\quad\quad\quad\quad\quad}
		\label{fig:element_groups}
	\end{subfigure}
	\caption{Concept of the proposed method for the patch-wise quadrature of trimmed surfaces illustrated by the example of the infinite plate with a circular hole. The mesh consists of $8\times8$ elements and the polynomial degrees are $p=q=2$. The trimming curve is illustrated by a red line.}
	\label{fig:concept}
\end{figure}

Fig. \ref{fig:element_numbering} shows the distinction in \textit{active-untrimmed}, \textit{trimmed} and \textit{inactive} elements as described in Section \ref{sec:integration_trimmed}. The basis functions can be also distinguished in the same groups based on the path of the trimming curve as illustrated in Fig. \ref{fig:basis_function_distinction}. Trimmed basis functions mean functions which span into trimmed elements and, therefore, are cut. On the other hand, untrimmed basis functions indicate functions which do not span into any trimmed element and, therefore, are not cut. The key idea is now to integrate the trimmed functions differently than the untrimmed ones. The patch-wise quadrature rule is still applicable to the latter. On the other hand, the trimmed basis functions can be integrated by a specialized quadrature rule for trimmed elements as described in Section \ref{sec:integration_trimmed}. Since the patch-wise quadrature rule does not integrate element-wise as outlined in Section \ref{sec:patch_wise_integration}, special care is required in the transition zone where untrimmed as well as trimmed basis functions occur in the same element (see Fig. \ref{fig:basis_function_support}). The elements are distinguished in four groups as visualized in Fig. \ref{fig:element_groups} (abbreviations for the groups are provided in brackets) and the applied quadrature rules are determined as:
\begin{itemize}
	\item \textit{Inactive} (\textit{ia}) $\Longrightarrow$ not integrated
	\item \textit{Trimmed} (\textit{t}) $\Longrightarrow$ mapped Gaussian quadrature rule (see Section \ref{sec:integration_trimmed})
	\item \textit{Transition} (\textit{tra}) $\Longrightarrow$ mixed integration (see below)
	\item \textit{Patch-wise} (\textit{pw}) $\Longrightarrow$ patch-wise quadrature rule (see Section \ref{sec:patch_wise_integration})
\end{itemize}

The distinction is similar to the one made in the general trimming case (compare Section \ref{sec:integration_trimmed}). However, the \textit{active-untrimmed} domain is further subdivided in a \textit{patch-wise} and a \textit{transition} domain. The \textit{inactive} elements are not integrated at all because they lie in the invisible part of the patch. The \textit{trimmed} elements are integrated by means of a mapped Gauss rule. We want to highlight that our proposed method is independent of the method used for the integration of the \textit{trimmed} elements and that any other quadrature rule could be applied equally well in our method. The \textit{patch-wise} elements are integrated by a patch-wise quadrature rule derived for the untrimmed patch. \bigskip 

The \textit{transition} elements require special care. They define elements where trimmed as well as untrimmed basis functions occur. The trimmed basis function can also have support in \textit{active-untrimmed} elements (see for example basis function 27 in Fig. \ref{fig:basis_function_support}). The size of the transition zone depends on the support of the trimmed basis functions and, therefore, on the degree $p$ since each basis function spans across ``$p+1$'' elements. \textit{Transition} elements are partly integrated by patch-wise quadrature points and partly by standard Gaussian quadrature points. Therefore, the term mixed integration is used for these elements. The different contributions in the \textit{transition} elements are integrated as follows:
\begin{itemize}
	\item Untrimmed test and untrimmed trial function $\Longrightarrow$ patch-wise quadrature rule (see Section \ref{sec:patch_wise_integration})
	\item Untrimmed test and trimmed trial function (or inversely) $\Longrightarrow$ Gauss rule (see Section \ref{sec:numerical_quadrature})
	\item Trimmed test and trimmed trial function $\Longrightarrow$ Gauss rule (see Section \ref{sec:numerical_quadrature})
\end{itemize}
The combined entries only appear in the transition zone. One should be aware that it is not possible to integrate the combined entries with the patch-wise quadrature rule because this rule is incapable of integrating contributions within element bounds. It is highlighted that the transition zone is unavoidable in order to link the patch-wise quadrature with trimmed elements. It is clear that the \textit{transition} elements require more computational effort than the other elements. The integrals, which should be computed, have the form of Eq.~\eqref{equ:integral_2D} or~\eqref{equ:integral_KL}. Thus, it is in general possible to select a different quadrature rule for each of the integrals belonging to single stiffness matrix entries.

\subsection{Efficiency Considerations}\label{sec:efficiency_considerations}
\begin{figure} [!bth]
	\centering
	\makebox[\textwidth][c]{\includegraphics[width=1.2\textwidth]{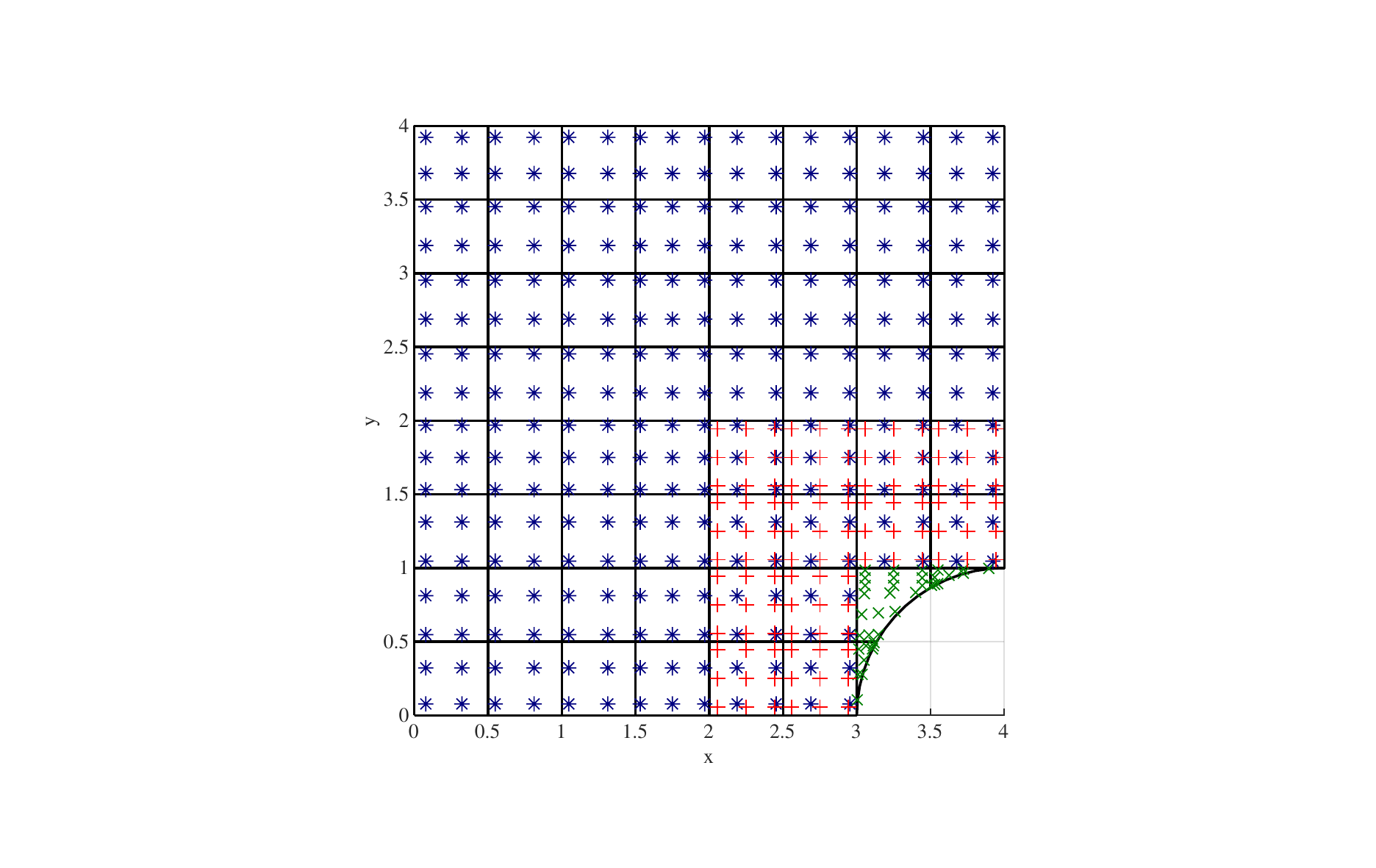}}
	\caption{Resulting quadrature points for the example of the infinite plate with circular hole given in Fig. \ref{fig:infinite_plate_with_circle} with $p=q=2$ and $8\times8$ elements. Stars in dark blue = patch-wise quadrature points, plus signs in \reviewed{red} = Gauss points in \textit{transition} elements, crosses in green = mapped Gauss points in \textit{trimmed} elements}
	\label{fig:pw_trimmed_integration_points}
\end{figure}

A reduction of quadrature points due to the use of the presented method is beneficial with respect to the computational cost. The setup of the patch-wise quadrature rule and the extra considerations due to the proposed method increase the effort in the preprocessing. However, this preprocessing time is relatively low in comparison to the total computational time. This is even more valid when the computations at the single quadrature points are often repeated - such as in dynamic as well as in non-linear analysis. An easy and well comparable possibility of assessing the improvement of efficiency is to count the number, and such the reduction, of the quadrature points. The patch-wise as well as the Gauss quadrature points are counted in the \textit{transition} elements whenever a total number of quadrature points is provided hereafter. A more involved implementation would possibly enable that only the necessary contributions to the element stiffness matrix are computed. Nevertheless, this approach of counting can be understood as upper bound for the efficiency assessment. Fig. \ref{fig:pw_trimmed_integration_points} exemplarily illustrates all the quadrature points involved in our proposed method for the discussed example. The expenses are clearly increased in all \textit{transition} elements compared to a pure Gauss quadrature (see Fig. \ref{fig:pw_trimmed_integration_points}). \bigskip

Coarse meshes may result in an even increased number of points compared to a standard Gauss integration under certain circumstances. This depends on the number of \textit{patch-wise} compared to \textit{transition} elements. \reviewed{The method proposed by \cite{Messmer2022} and the Discontinuous Weighted Quadrature proposed by \cite{Marussig2022} are able to directly reduce the number of quadrature points also for coarse meshes - this does not hold for the Hybrid Gauss Approach presented in \cite{Marussig2022}. However, they seem to require a comparable number of quadrature points for finer meshes, which are needed for accurate results. A direct comparison is not possible because Meßmer et al. \cite{Messmer2022} treat volumes and Marussig \cite{Marussig2022} uses Weighted Quadrature. For the proposed method,} it can be estimated in advance whether an application reduces the number of quadrature points for a specific trimmed patch. If a Gauss rule is used, the number of quadrature points is:
\begin{equation}\label{equ:n_gauss}
	n_{gauss} = (p+1)\cdot (q+1)\cdot n_{ele}
\end{equation}
If the patch-wise quadrature rule is applied in case of the plane element, the number of integration points can be estimated applying Eq.~\eqref{equ:n_quad_2D} as: 
\begin{equation}\label{equ:n_pw-trimm_2D}
	n_{pw-trimm,plane} = (p+1)\cdot (q+1)\cdot (n_{ele,t}+n_{ele,tra})+((p+2)\cdot (q+2))/4\cdot (n_{ele,pw}+n_{ele,tra})
\end{equation}
where the term $pw-trimm$ in the index indicates the proposed method, $n_{ele,t}$ is the number of \textit{trimmed} elements, $n_{ele,tra}$ is the number of \textit{transition} elements and $n_{ele,pw}$ is the number of \textit{patch-wise} elements. If the patch-wise quadrature rule is applied in case of the Kirchhoff-Love shell element, the number of quadrature points can be estimated applying Eqs.~\eqref{equ:n_quad_KL} as:
\begin{equation}\label{equ:n_pw-trimm_KL}
	n_{pw-trimm,KL} = (p+1)\cdot (q+1)\cdot (n_{ele,t}+n_{ele,tra})+((p+3)\cdot (q+3))/4\cdot (n_{ele,pw}+n_{ele,tra})
\end{equation}
Eqs.~\eqref{equ:n_pw-trimm_2D} and~\eqref{equ:n_pw-trimm_KL} consider in the first term that \textit{trimmed} and \textit{transition} elements contain ``$p$+1'' Gauss points per direction and in the second term that the \textit{patch-wise} and the \textit{transition} elements contain patch-wise quadrature points. These equations are only approximate and assume maximum regularity (see Section \ref{sec:patch_wise_integration}). Considering Eqs.~\eqref{equ:n_gauss},~\eqref{equ:n_pw-trimm_2D} and~\eqref{equ:n_pw-trimm_KL}, the proposed method should be only used, if the following condition is fulfilled:
\begin{equation}\label{equ:use_rule}
	n_{pw-trimm} < n_{gauss}
\end{equation}
where $n_{pw-trimm}$ is either obtained by Eq.~\eqref{equ:n_pw-trimm_2D} or~\eqref{equ:n_pw-trimm_KL} in case of plane or KL shell elements, respectively. \reviewed{The same quadrature rule for \textit{trimmed} elements in case of standard Gauss and the proposed patch-wise quadrature is used. Therefore, \textit{trimmed} elements can be excluded from Eq.~\eqref{equ:use_rule} which leads to a comparison of the ratio of \textit{transition} and \textit{patch-wise} elements:}
\begin{equation}\label{equ:use_rule_a}
	\dfrac{n_{ele,tra}}{n_{ele,pw}} < c
\end{equation}
\reviewed{where $c$ is a constant depending on the applied element formulation. In case of the plane element, $c$ is defined as:}
\begin{equation}\label{equ:constant_c_plane}
	c = \dfrac{4\cdot(p+1)\cdot(q+1)-(p+2)\cdot(q+2)}{(p+2)\cdot(q+2)}
\end{equation}
\reviewed{In case of the KL shell element, $c$ is defined as:}
\begin{equation}\label{equ:constant_c_KL}
	c = \dfrac{4\cdot(p+1)\cdot(q+1)-(p+3)\cdot(q+3)}{(p+3)\cdot(q+3)}
\end{equation}

It is highlighted that there is always a clear reduction of quadrature points with the proposed method with further refinement as demonstrated in the following numerical examples.

\section{Numerical Results}\label{sec:numerical_results}
In this section, multiple numerical examples are investigated in order to assess the proposed method (denoted in this section as \textit{pw-trimm}). The infinite plate with a circular hole, which was already used in Section \ref{sec:patch_wise_integration_trimmed_surface} for the explanation of the method, is discussed as a first example. This problem is a plane strain model. The following two examples in Section \ref{sec:plate} and \ref{sec:punched_plate} are plate problems. Subsequently, the method is tested for a shell benchmark in Section \ref{sec:scordelis}. The application of the method in the industrial context is demonstrated by the last example of a tube connection. A Kirchhoff-Love shell element formulation, which was firstly introduced by \cite{Kiendl2011}, is used for the plate as well as the shell examples. It is highlighted that the proposed method is independent of the element formulation, but the patch-wise quadrature rule itself has to consider it as discussed in Section \ref{sec:patch_wise_integration}. All models are computed with linear elastic materials. Geometrically linear as well as non-linear computations are performed. An efficient quadrature rule is especially interesting for non-linear analyses where computations at the quadrature points are iteratively repeated many times. A diagonal scaling preconditioner is employed in order to avoid ill-conditioning of the system of equations, which might occur due to very small trimmed elements. A detailed description of this preconditioner is provided by \cite{Antolin2019}. The results with the new method (\textit{pw-trimm}) are compared to results obtained from a standard trimming procedure which is denoted as \textit{trimm} in this section. In addition to the results reported in the following sections, single displacements are documented as reference in \ref{app:displacement_results} to facilitate the reproducibility of the examples.

\subsection{Infinite Plate with Circular Hole}\label{sec:infinite_plate}
The setup of the infinite plate with a circular hole is illustrated in Fig. \ref{fig:infinite_plate_with_circle}. Slightly different setups are found in the literature with respect to the actual loading and the Young's modulus \cite{Hughes2005,Antolin2019,Coradello2020b}. All computations are performed on the depicted trimmed geometry. Only a quarter of the infinite plate is modelled by making use of the symmetry and applying the exact traction at the left and upper boundaries as Neumann conditions:
\begin{align}
	\sigma_{rr}(r,\theta) &= \dfrac{T_x}{2} \left(1-\dfrac{R^2}{r^2}\right) + \dfrac{T_x}{2} \left(1-4\dfrac{R^2}{r^2}+3\dfrac{R^4}{r^4}\right) \cos(2\theta) \\
	\sigma_{\theta\theta}(r,\theta) &= \dfrac{T_x}{2} \left(1+\dfrac{R^2}{r^2}\right)-\dfrac{T_x}{2} \left(1+3\dfrac{R^4}{r^4}\right) \cos(2\theta) \\
	\sigma_{r\theta}(r,\theta) &= -\dfrac{T_x}{2} \left(1+2\dfrac{R^2}{r^2}-3\dfrac{R^4}{r^4}\right) \sin(2\theta)
\end{align}
where the parameters are illustrated in Fig. \ref{fig:infinite_plate_with_circle}. \bigskip

\begin{figure} [!bth]
	\centering
	\includegraphics{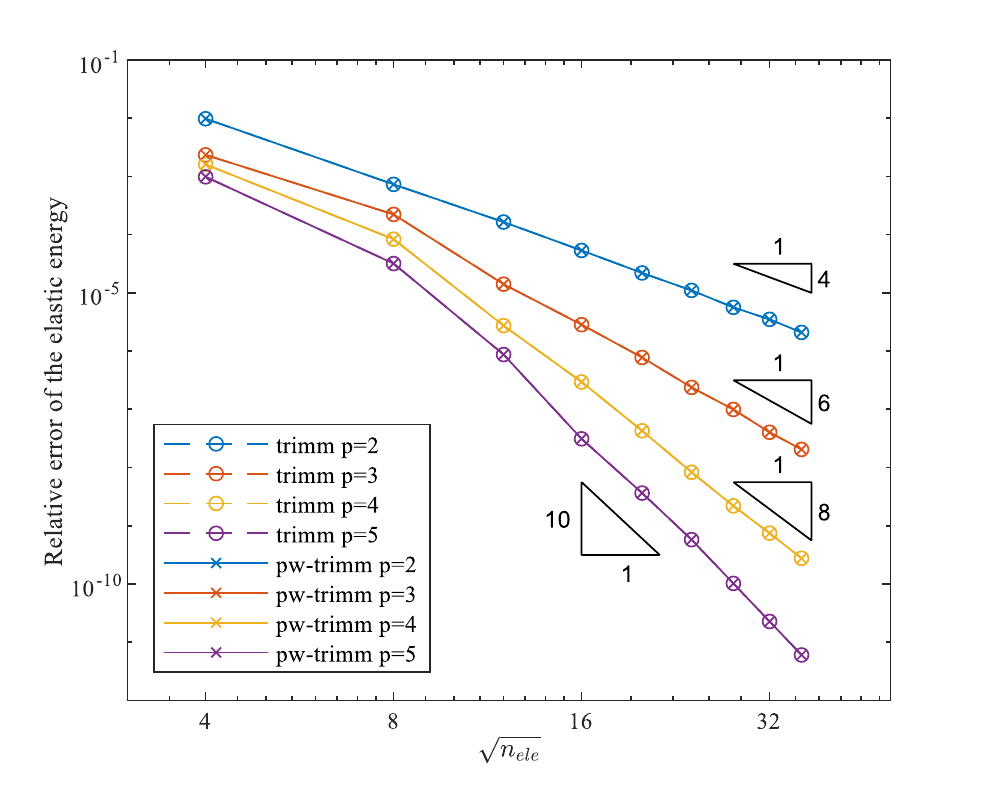}
	\caption{Convergence plot of the relative error of the elastic energy for the infinite plate with circular hole.}
	\label{fig:infinite_plate_energy}
\end{figure}

A convergence study of the relative error of the elastic energy, $error=|W-W_{ref}|/W_{ref}$, is performed. The elastic energy is computed as $W = \frac{1}{2}\int_\Omega (\mathbf{\sigma}:\mathbf{\epsilon}) d\Omega$. The analytical reference solution is $W_{ref}=7.69365373$. Basis functions with different polynomial degrees from $p=2$ to $p=5$ are studied. The convergence curves are illustrated in Fig. \ref{fig:infinite_plate_energy}. The results from \textit{trimm} and \textit{pw-trimm} exactly match. The highest relative difference is around $10^{-14}$ which can be seen as a purely numerical round-off error. Optimal convergence rates for all polynomial degrees are observed. \bigskip

\begin{figure} [!bth]
	\centering
	\includegraphics{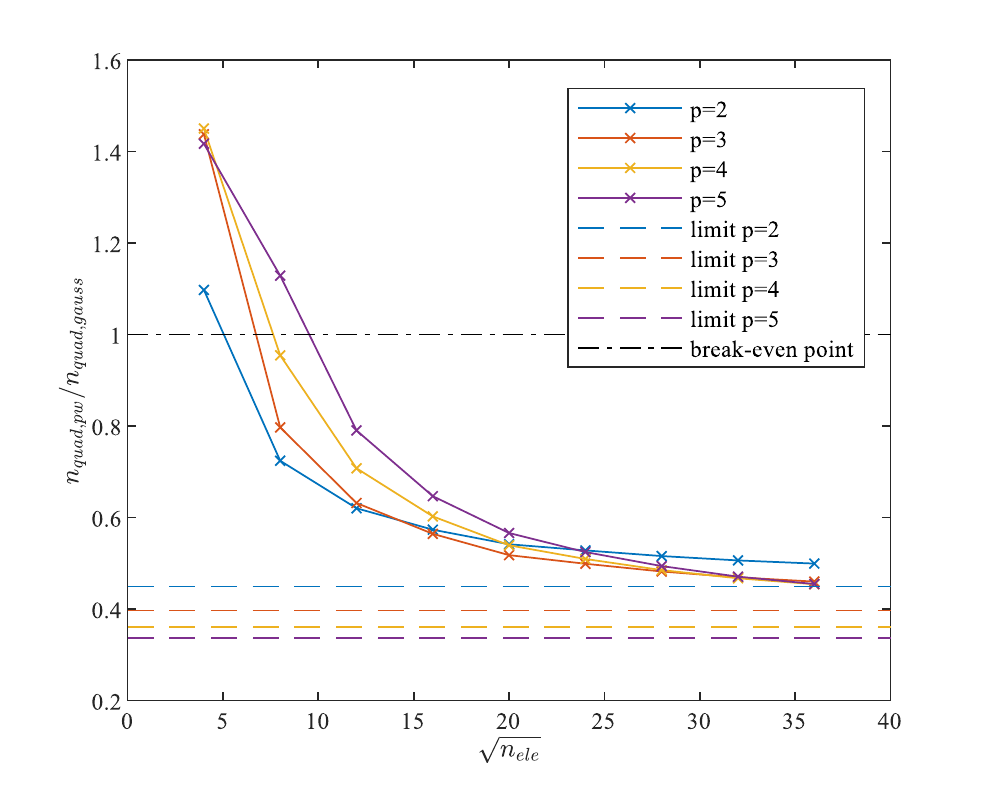}
	\caption{Ratio of the number of quadrature points of the patch-wise quadrature $n_{quad,pw}$ versus the Gaussian quadrature $n_{quad,gauss}$ for the infinite plate with circular hole. The theoretical limits of this ratio are provided as dashed lines.}
	\label{fig:infinite_plate_reduction_GP}
\end{figure}

\begin{figure} [!bth]	
	\centering
	\begin{subfigure}{0.49\textwidth}
		\centering
		\includegraphics{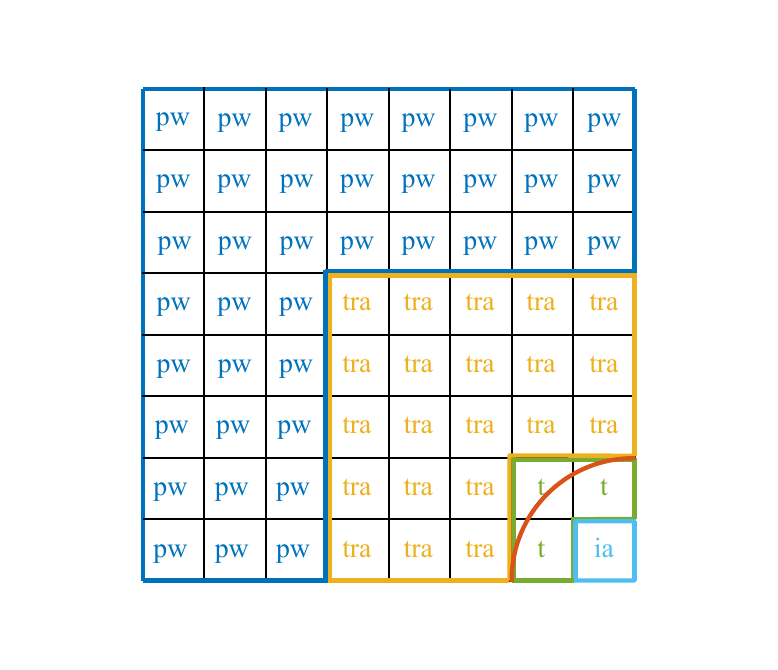}
		\caption{Element distinction for degree $p=3$.}
		\label{fig:element_groups_p3}
	\end{subfigure}
	\begin{subfigure}{0.49\textwidth}
		\centering	
		\includegraphics{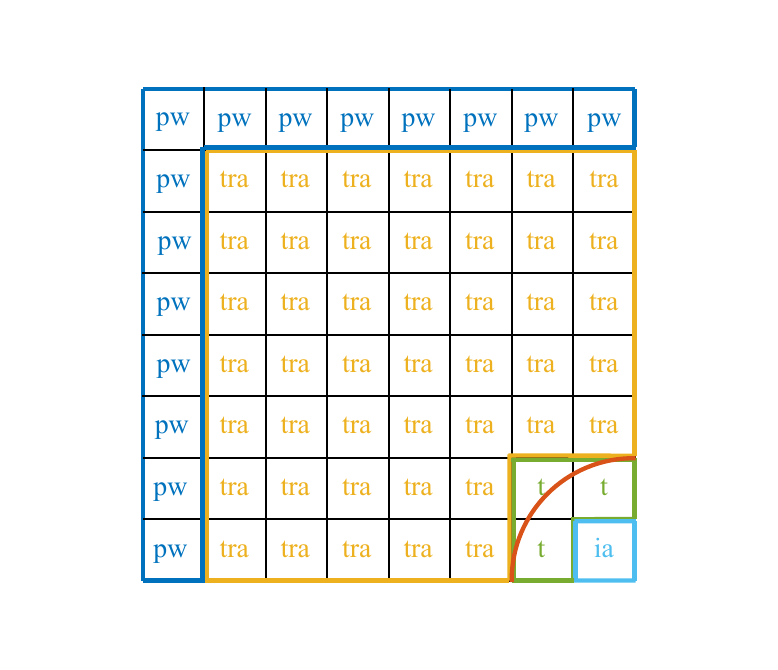}
		\caption{Element distinction for degree $p=5$.}
		\label{fig:element_groups_p5}
	\end{subfigure}
	\caption{Comparison of element distinctions in the proposed method for the polynomial degrees $p=3$ and $p=5$ indicated by colors and abbreviations. \textit{pw} in dark blue = \textit{patch-wise}, \textit{tra} in gold = \textit{transition}, \textit{t} in green = \textit{trimmed}, \textit{ia} in light blue = \textit{inactive}. Trimming curve in red.}
	\label{fig:element_groups_comparison}
\end{figure}

The reduction of quadrature points by means of the proposed method is illustrated in Fig. \ref{fig:infinite_plate_reduction_GP}. In case of polynomial degrees up to $4$, a reduction is already observed with the second mesh with $8\times8$ elements. In case of a degree $p=5$, a reduction is achieved with the third mesh with $12\times12$ elements. The models with higher polynomial degrees show less gains for coarse meshes but higher reduction for fine meshes. Basis functions with a higher degree have support over more elements resulting in more \textit{transition} elements as illustrated in Fig. \ref{fig:element_groups_comparison}. On the other hand, patch-wise rules are more effective for higher degrees. This means that the reduced number of integration points in the \textit{patch-wise} elements compensates at a certain point for the higher number of \textit{transition} elements. Furthermore, the theoretical limits which could be obtained by infinitely fine meshes are illustrated in Fig.  \ref{fig:infinite_plate_reduction_GP}. These limits are obtained from the 1D-limits documented in Table \ref{tab:reduction_n_quad}. They illustrate that the proposed method asymptotically approaches the ideal number of quadrature points due to a patch-wise quadrature rule.

\begin{figure} [!bth]
	\centering
	\includegraphics{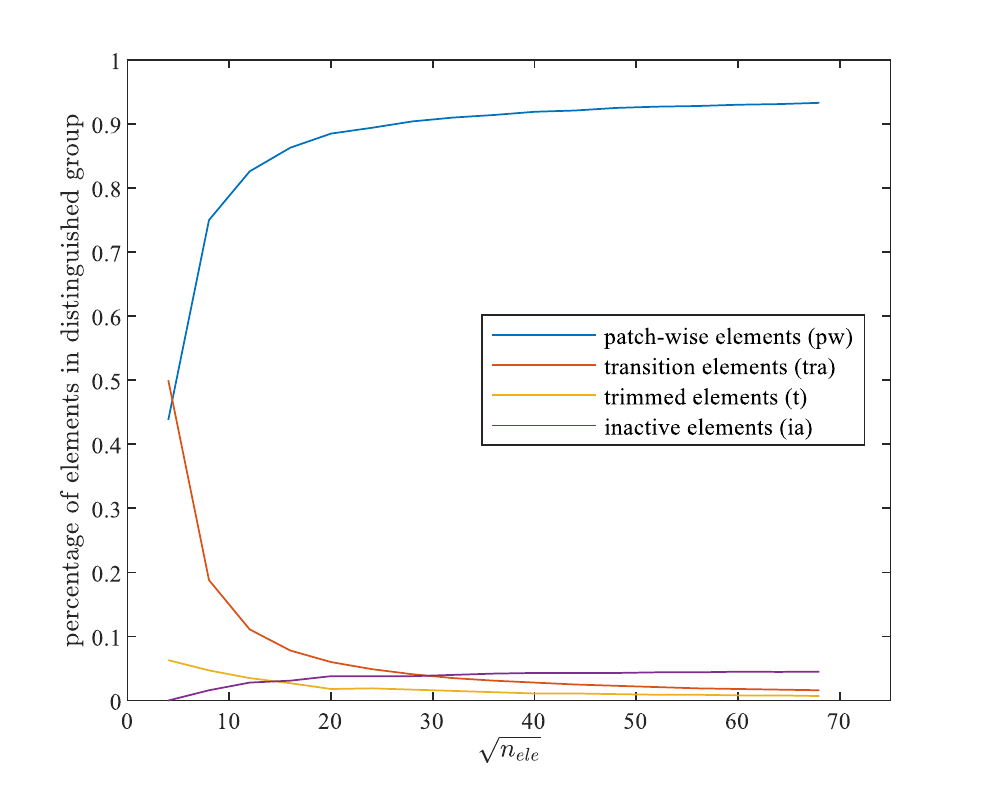}
	\caption{\reviewed{Percentage of elements in a distinguished group (see Section \ref{sec:concept}) with respect to the total number of elements for the infinite plate with circular hole with $p=2$.}}
	\label{fig:infinite_plate_n_ele}
\end{figure}

\reviewed{Fig. \ref{fig:infinite_plate_n_ele} shows the percentage distribution of the elements in the distinguished groups (namely \textit{patch-wise}, \textit{transition}, \textit{trimmed}, \textit{inactive}) for $p=2$. In contrast to the previous figures, finer meshes are additionally depicted to reveal the asymptotic behaviour. It is observed that the percentage of \textit{patch-wise} elements clearly dominates for refined meshes. The proportion of \textit{transition} and \textit{trimmed} continuously decreases and almost vanishes. The percentage of \textit{inactive} elements with $4.50\%$ for the finest depicted mesh asymptotically reaches the proportion of the hole with respect to the complete area which is $4.91\%$.}

\subsection{Plate with circular hole}\label{sec:plate}
\begin{figure} [!bth]
	\centering
	\includegraphics{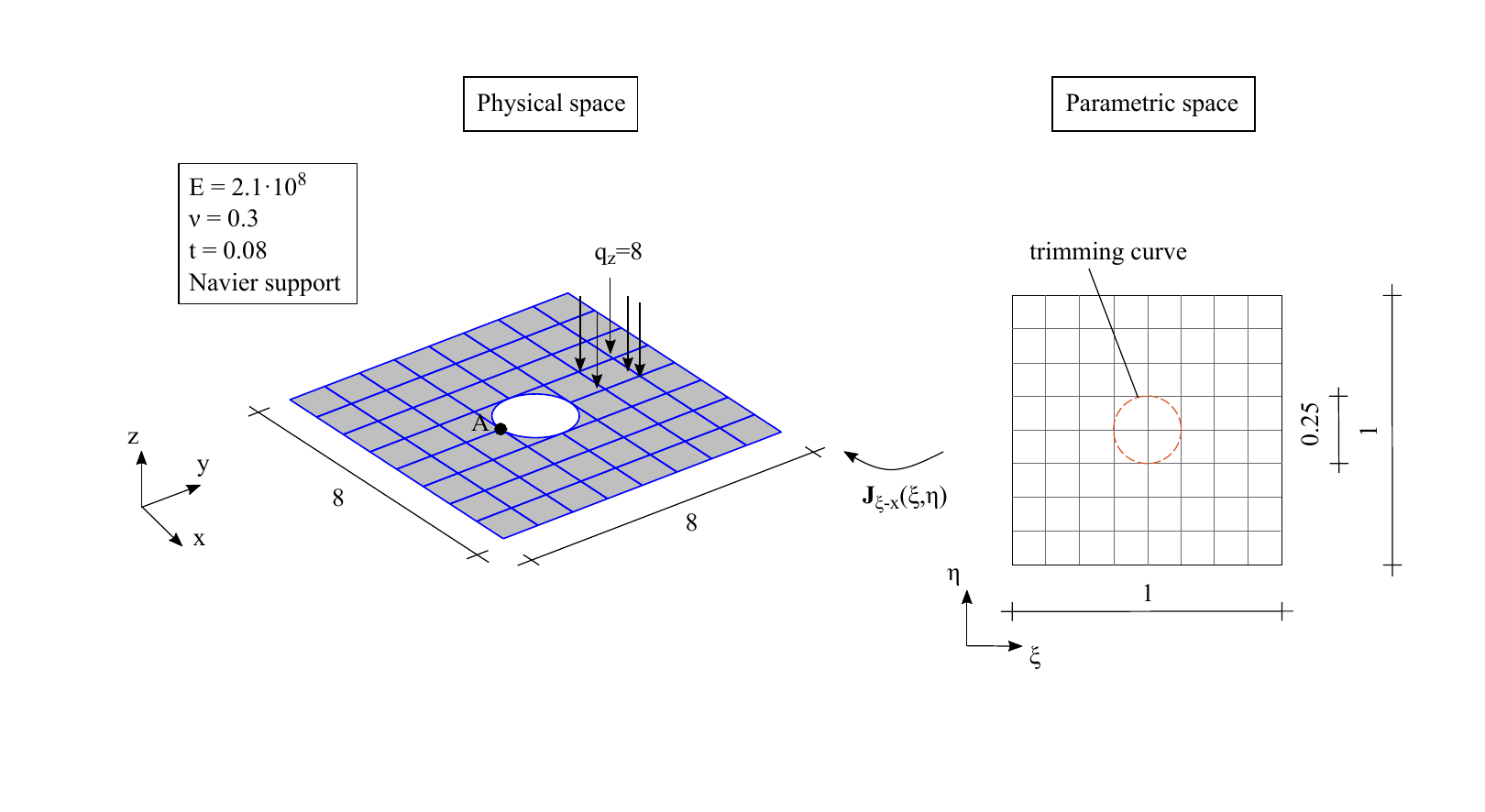}
	\caption{Setup of the plate with circular hole.}
	\label{fig:plate_system}
\end{figure}

\begin{figure} [!bth]
	\centering
	\includegraphics{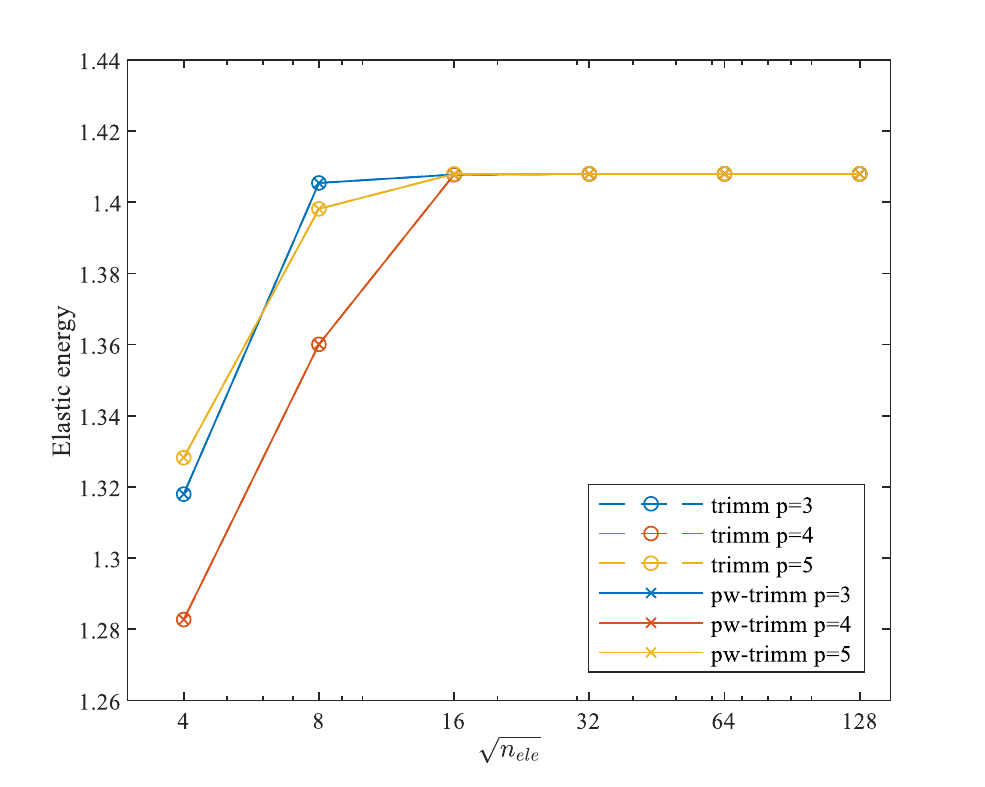}
	\caption{Convergence plot of the elastic energy for the plate with circular hole.}
	\label{fig:plate_energy}
\end{figure}

The setup of the plate with a circular hole is illustrated in Fig. \ref{fig:plate_system}. The geometry is in principle the same as in Section \ref{sec:infinite_plate}, but modelled as complete plate without considering symmetry conditions. The plate has Navier supports (supports in $z$-direction) at all edges. A convergence study of the elastic energy is performed studying different polynomial degrees from $p=3$ to $p=5$. The convergence curves are reported in Fig. \ref{fig:plate_energy}. The results from \textit{trimm} and \textit{pw-trimm} again match exactly. The highest absolute difference is around $10^{-10}$. The result of the elastic energy converges with a precision of eight valid digits to a value of $W = 1.4079595$ for the illustrated meshes. \bigskip

\begin{figure} [!bth]
	\centering
	\includegraphics{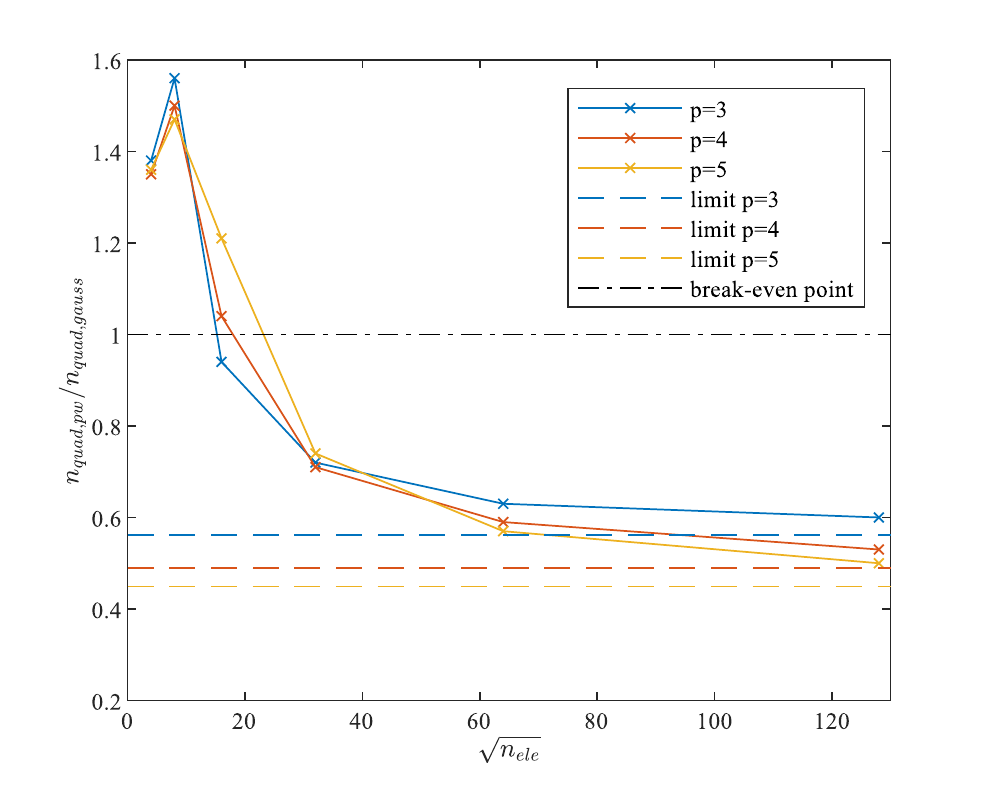}
	\caption{Ratio of the number of quadrature points of the patch-wise quadrature $n_{quad,pw}$ versus the Gaussian quadrature $n_{quad,gauss}$ for the plate with circular hole. The theoretical limits of this ratio are provided as dashed lines.}
	\label{fig:plate_reduction_GP}
\end{figure}

The reduction of quadrature points by means of the proposed method is illustrated in Fig. \ref{fig:plate_reduction_GP}. A reduction is observed for all degrees with the fourth mesh with $32\times32$ elements (for $p=3$ already with the third mesh). For coarse meshes, the proposed method actually requires more quadrature points than Gaussian quadrature because the \textit{transition} elements, which have a higher number of points, dominate. It is reminded that Eq.~\eqref{equ:use_rule} provides an a-priori check of the expectable efficiency of the patch-wise rule. Furthermore, the counting of the number of quadrature points in the \textit{transition} elements can be seen as a conservative estimate with respect to the efficiency as described in Section \ref{sec:efficiency_considerations}. Nevertheless, it becomes clear, especially in comparison to the results from Fig. \ref{fig:infinite_plate_reduction_GP}, that the efficiency gains due to the proposed method highly depend on the considered geometry. A reduction of points can be always expected for finer meshes.

\subsection{Punched plate}\label{sec:punched_plate}

\begin{figure} [!bth]
	\centering
	\includegraphics{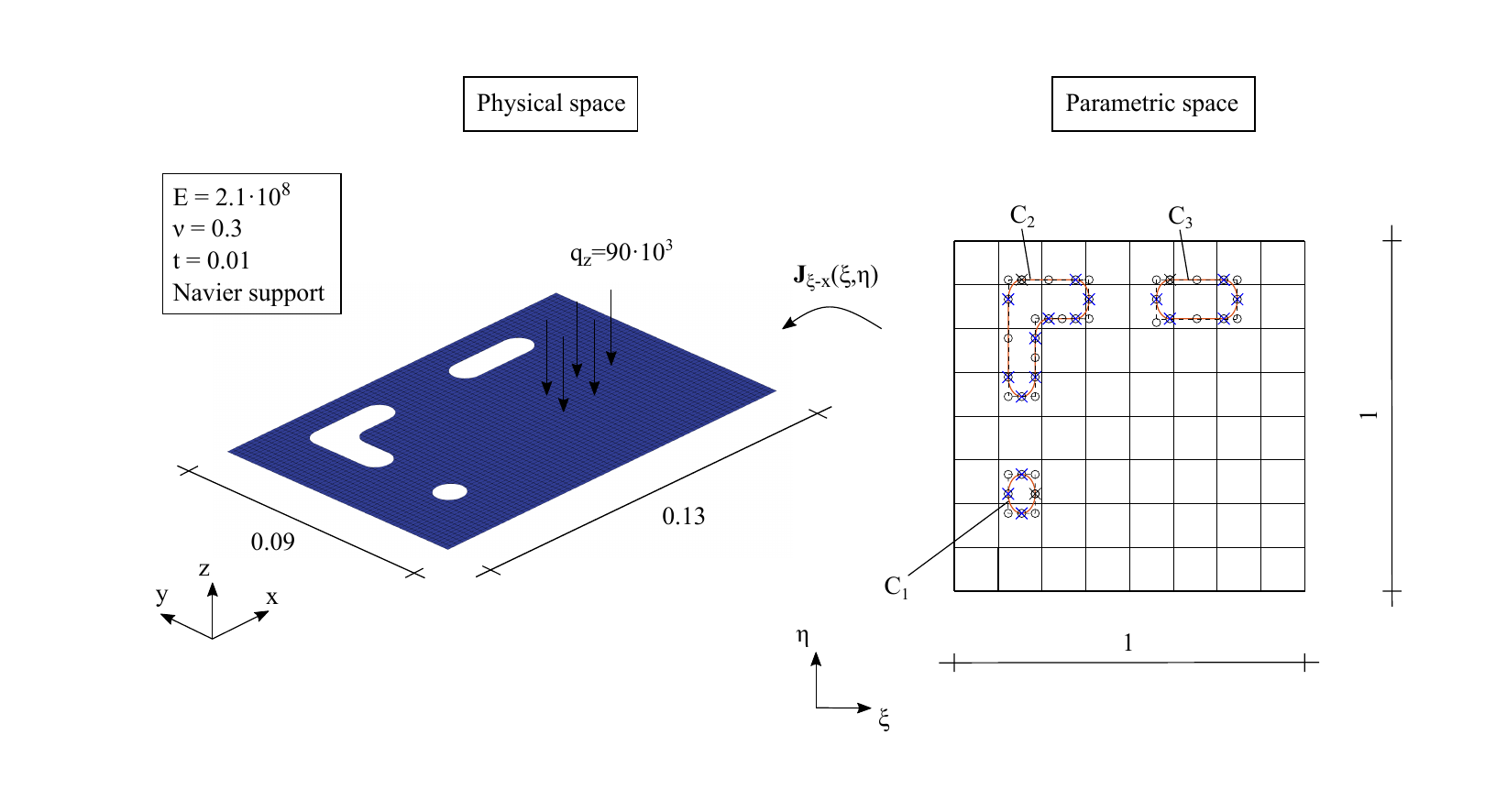}
	\caption{Setup of the punched plate. Description of trimming curves in the parametric space: red curve = trimming curve, black circles = control point, black cross = start and end point of curve, blue cross = knot. NURBS description of the trimming curves in \ref{app:punched_plate_nurbs}.}
	\label{fig:punched_plate_system}
\end{figure}

\begin{figure} [!bth]
	\centering
	\includegraphics{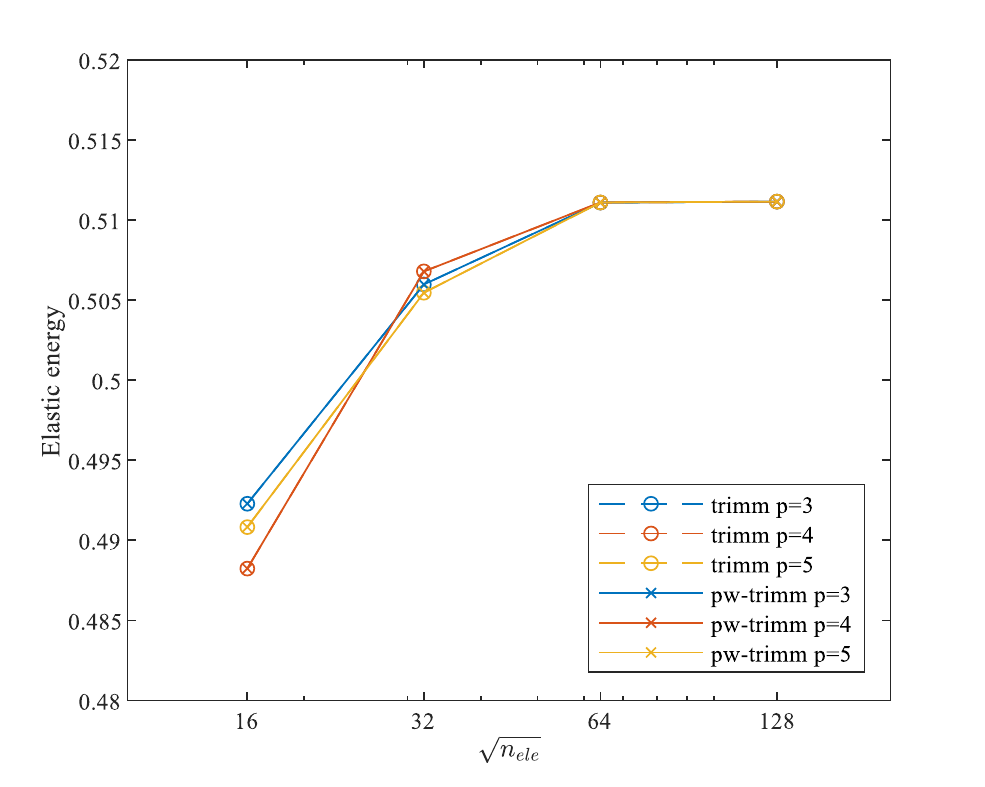}
	\caption{Convergence plot of the elastic energy for the punched plate.}
	\label{fig:punched_plate_energy}
\end{figure}

The following example of a punched plate investigates the performance of the presented method for multiple holes with more complex shapes. The setup is illustrated in Fig. \ref{fig:punched_plate_system}. The plate has Navier supports (supports in $z$-direction) at all edges. The exact NURBS description of the three trimming curves is provided in \ref{app:punched_plate_nurbs}. A convergence study of the elastic energy is performed and the results are illustrated in Fig. \ref{fig:punched_plate_energy}. Spurious coupling occurs for too coarse meshes due to basis functions spanning across the holes \cite{Coradello2020a}. This effect depends on the mesh size as well as on the size and position of the features (holes). The infinite plate with hole presented in Section \ref{sec:infinite_plate} for example does not suffer by this problem because its hole is positioned at the boundary. A recommendable solution to alleviate this shortcoming would be local refinement as discussed in \cite{Coradello2020a,Coradello2020b}. The results from \textit{trimm} and \textit{pw-trimm} again match exactly. The highest absolute difference is around $2\cdot 10^{-10}$.  The result of the elastic energy converges with a precision of four valid digits to a value of $W=0.5112$ for the illustrated meshes. \bigskip

\begin{figure} [!bth]
	\centering
	\includegraphics{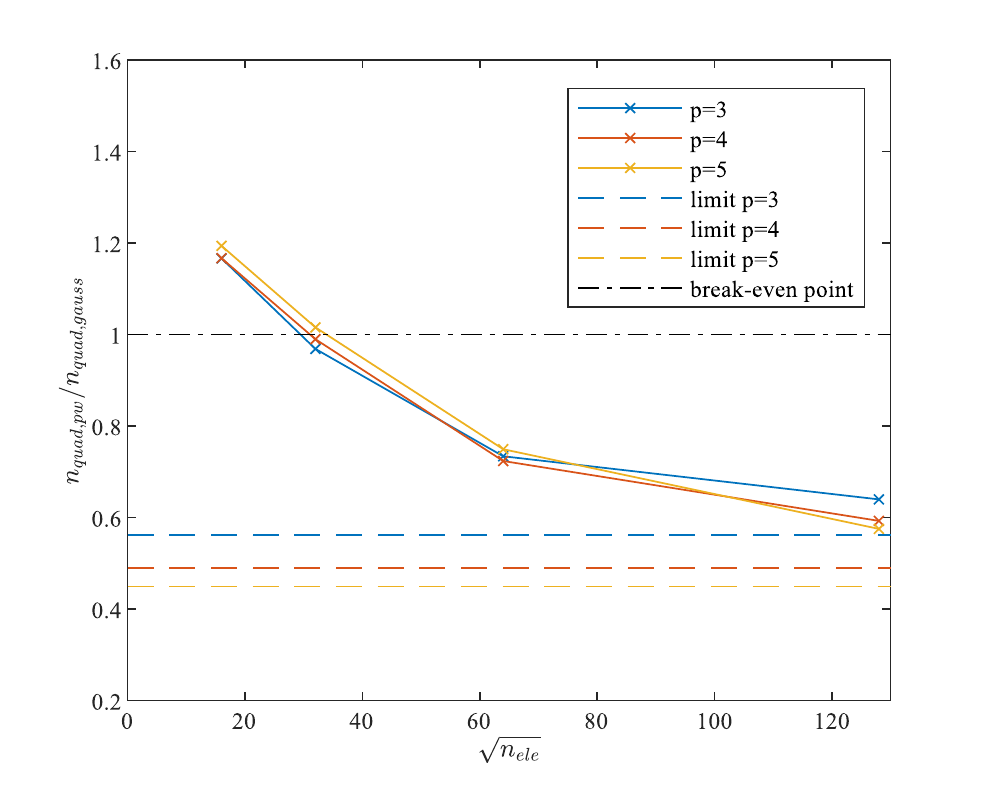}
	\caption{Ratio of the number of quadrature points of the patch-wise quadrature $n_{quad,pw}$ versus the Gaussian quadrature $n_{quad,gauss}$ for the punched plate. The theoretical limits of this ratio are provided as dashed lines.}
	\label{fig:punched_plate_reduction_GP}
\end{figure}

The reduction of quadrature points by means of the proposed method is illustrated in Fig. \ref{fig:punched_plate_reduction_GP}. A reduction is observed for degrees $p=3$ and $p=4$ with the second mesh with $32\times32$ elements (for $p=5$ with the third mesh).

\subsection{Geometrically Linear Scordelis-Lo Roof with Elliptic Hole in the Centre}\label{sec:scordelis}
\begin{figure} [!bth]
	\centering
	\includegraphics{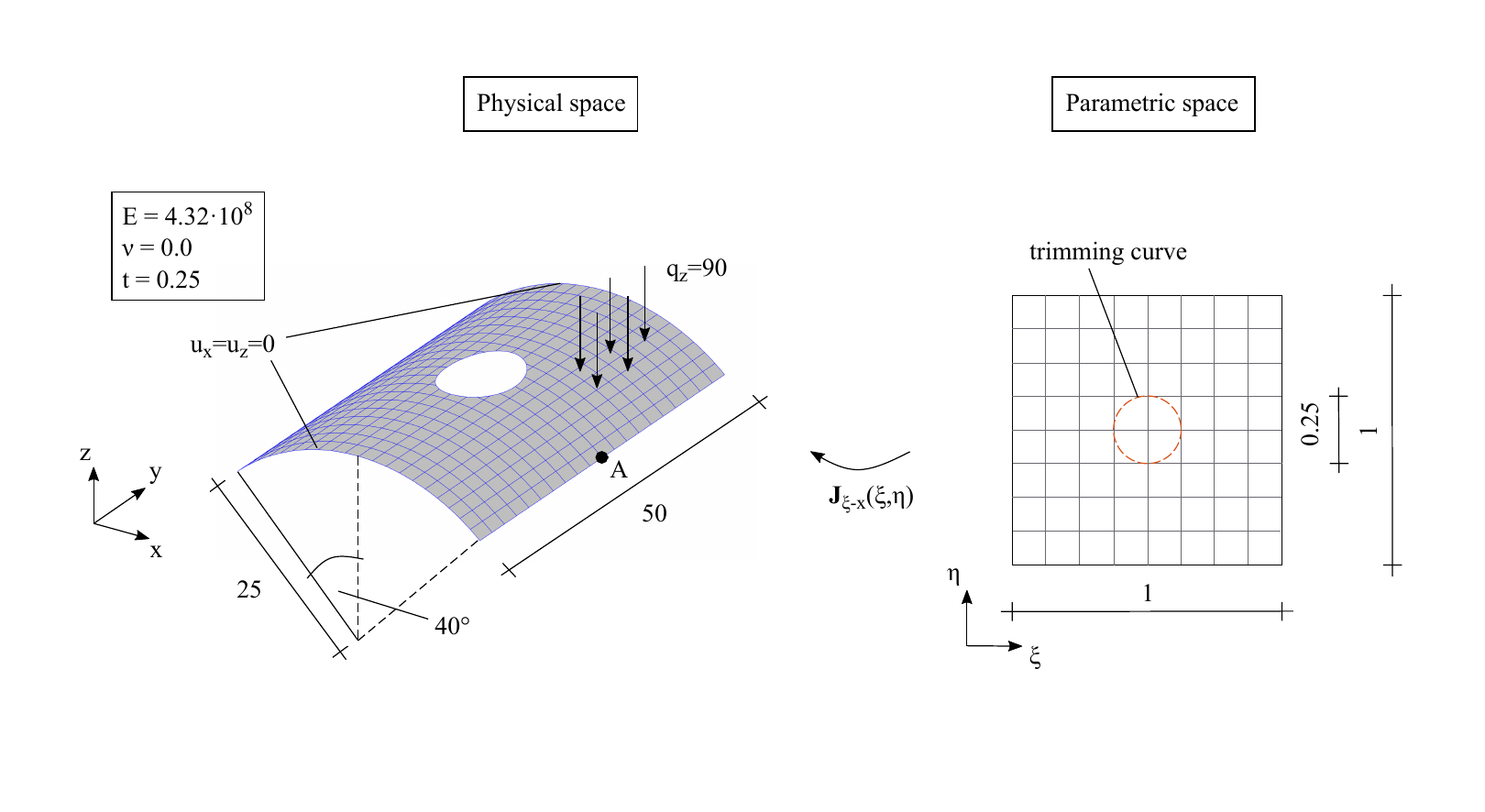}
	\caption{Setup of the Scordelis-Lo roof with elliptic hole.}
	\label{fig:scordelis_elliptic_hole}
\end{figure}

\begin{figure} [!bth]
	\centering
	\includegraphics{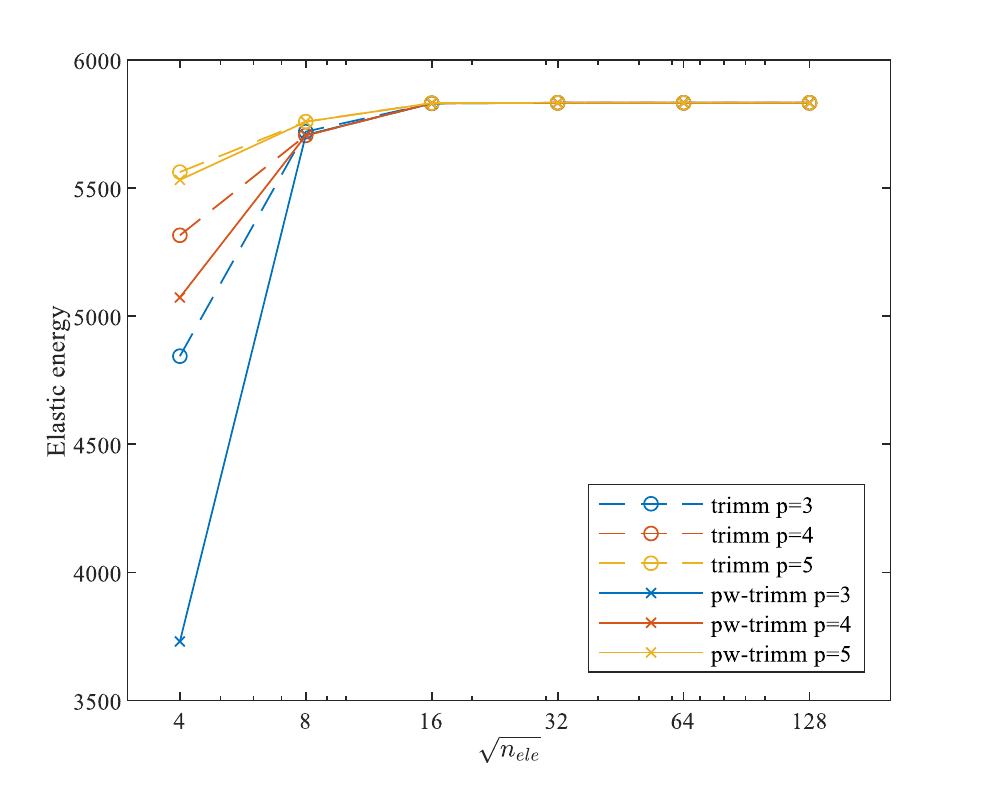}
	\caption{Convergence plot of the elastic energy for the Scordelis-Lo roof with elliptic hole. (For a better interpretation of the graphs due to used colours, the reader is referred to the web version of this article.)}
	\label{fig:scordelis_energy}
\end{figure}

The setup of the Scordelis-Lo roof with an elliptic hole in the centre is illustrated in Fig. \ref{fig:scordelis_elliptic_hole}. This shell model was initially proposed by \cite{Coradello2020b}. A geometrically linear analysis is performed. A convergence study of the elastic energy is carried out and is illustrated in Fig. \ref{fig:scordelis_energy}. The result of the elastic energy converges with a precision of seven valid digits to a value of $W=5832.471$ for the illustrated meshes. In contrast to the previous examples, a larger difference of the results from the standard trimming and the new patch-wise trimming procedure is observed which still reduces with mesh refinement. It is actually expected in this context that the results differ when using two different quadrature rules because both methods do not perfectly solve the considered integrals. This is in particular true for shells, like in this example, where the basis functions are really rational splines and the Jacobian is definitely non-constant. The \textit{trimm} results are more accurate especially for coarse meshes because they use a higher number of quadrature points which is beneficial for resolving the mentioned rational terms of the integrands. This interesting effect was not considered in literature so far. Even though, Hokkanen \cite{Hokkanen2020} also documented shortcomings of existing reduced patch-wise quadrature rules in the context of complex (real-world) shell applications. Therefore, a comparison of Gaussian and patch-wise quadrature in the context of highly curved shells is an interesting topic for future work. The reduction of quadrature points for this example is identical to the one for the plate with circular hole illustrated in Fig. \ref{fig:plate_reduction_GP} because the trimming patterns defined in the parametric space are identical (compare Fig. \ref{fig:plate_system} and \ref{fig:scordelis_elliptic_hole}). \bigskip

\subsection{Geometrically Non-Linear Scordelis-Lo Roof with Elliptic Hole}
\begin{figure} [!bth]
	\centering
	\includegraphics{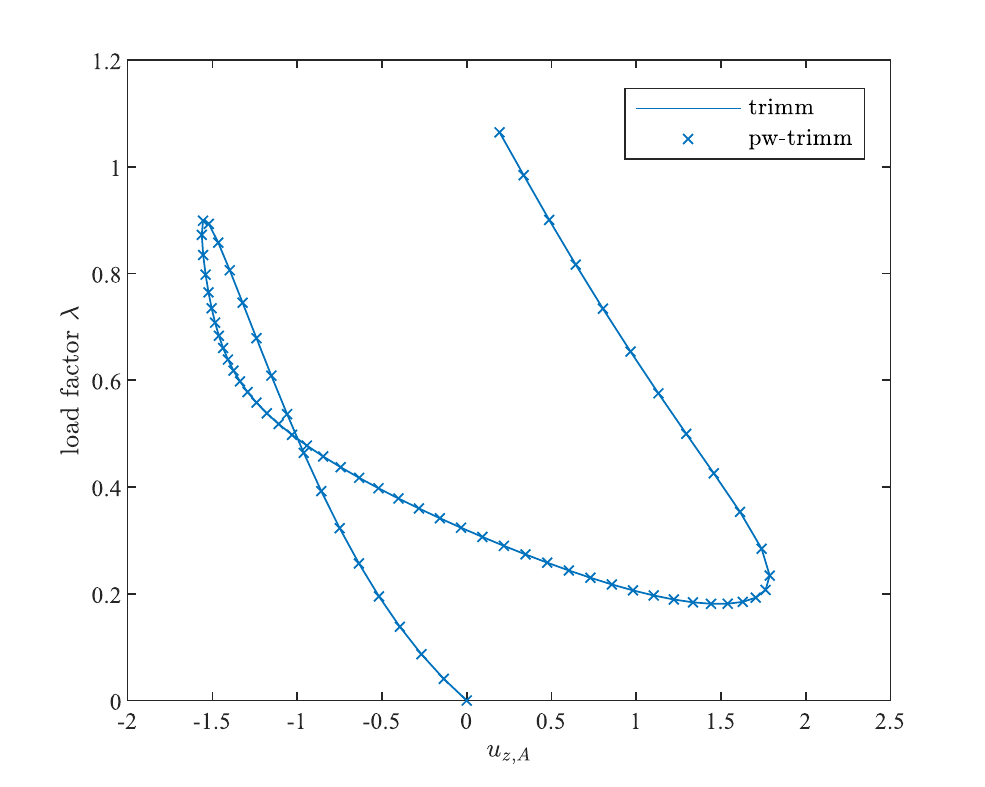}
	\caption{Load-displacement curve with load factor $\lambda$ and the displacement $u_{z,A}$ in z-direction at point A for the geometrically non-linear computed Scordelis-Lo roof with elliptic hole. The polynomial degree is $p=q=3$ and $64\times64$ elements are used.}
	\label{fig:scordelis_nonlin_load_displacement}
\end{figure}

\begin{figure} [!bth]	
	\centering
	\begin{subfigure}{0.49\textwidth}
		\centering
		\includegraphics[width=\textwidth,keepaspectratio]{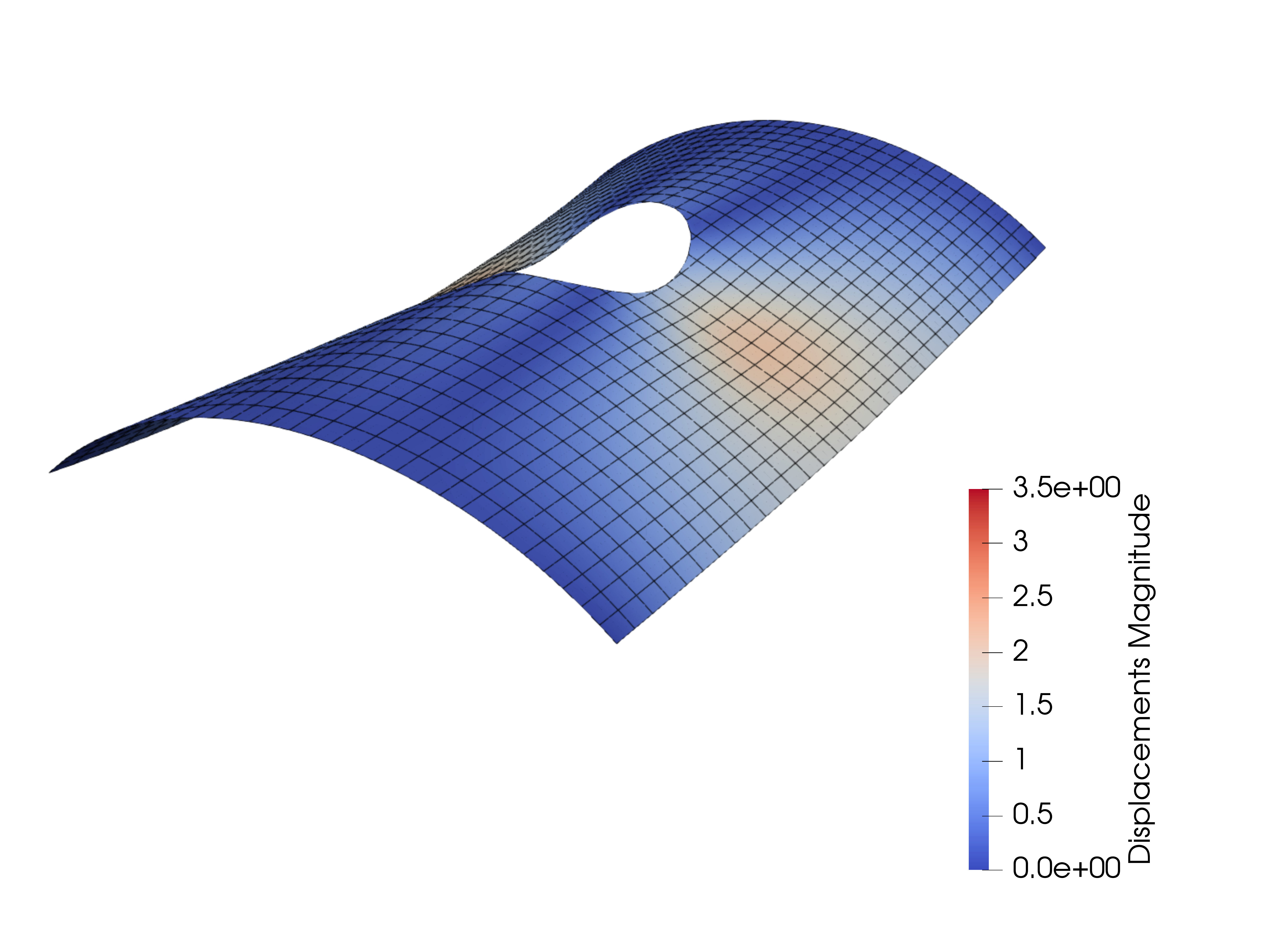}
		\caption{Load step 20.}
		\label{fig:scordelis_nonlin_step20}
	\end{subfigure}
	\begin{subfigure}{0.49\textwidth}
		\centering	
		\includegraphics[width=\textwidth,keepaspectratio]{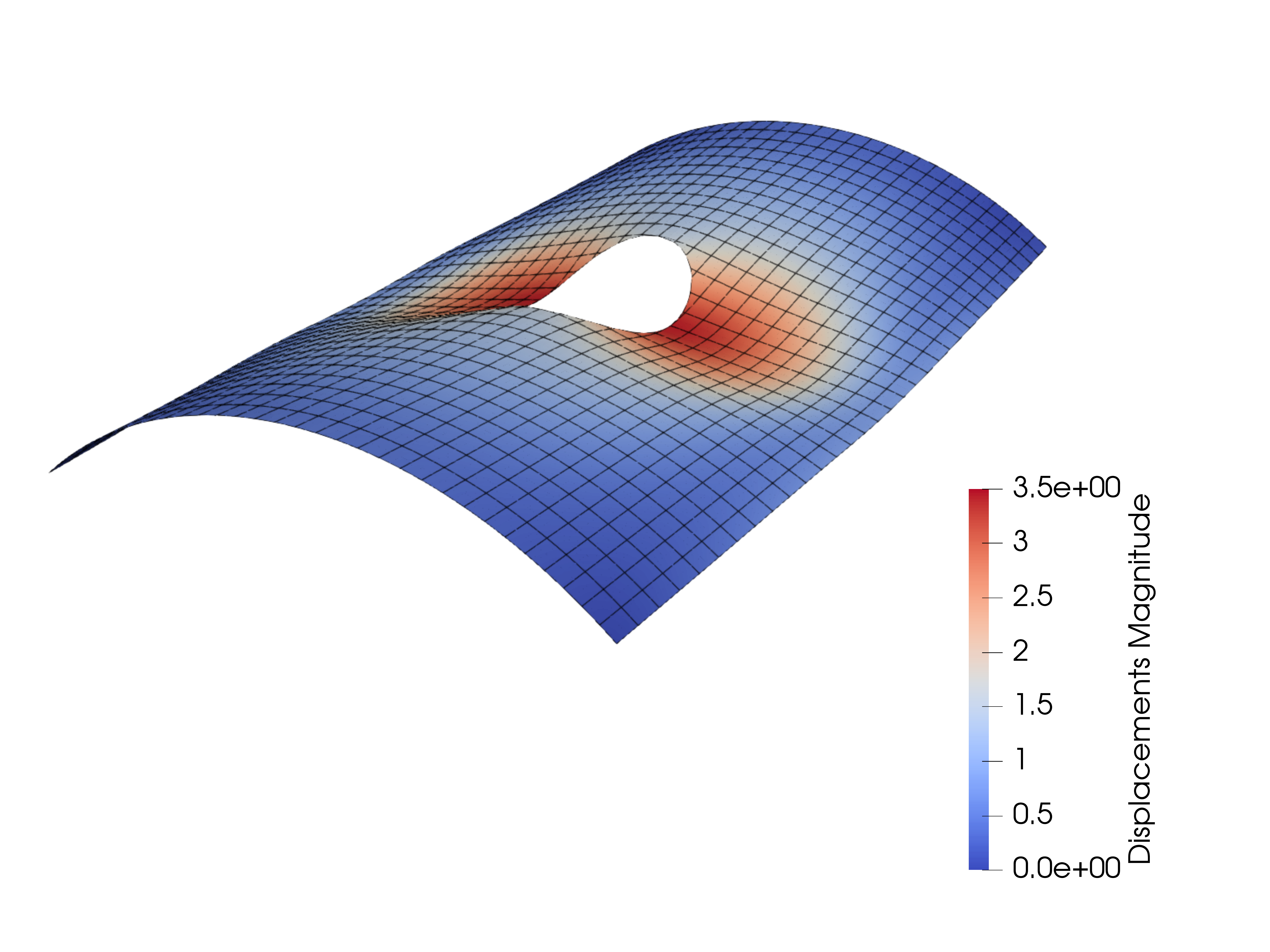}
		\caption{Load step 40.}
		\label{fig:scordelis_nonlin_step40}
	\end{subfigure}
	\begin{subfigure}{0.49\textwidth}
		\centering
		\includegraphics[width=\textwidth,keepaspectratio]{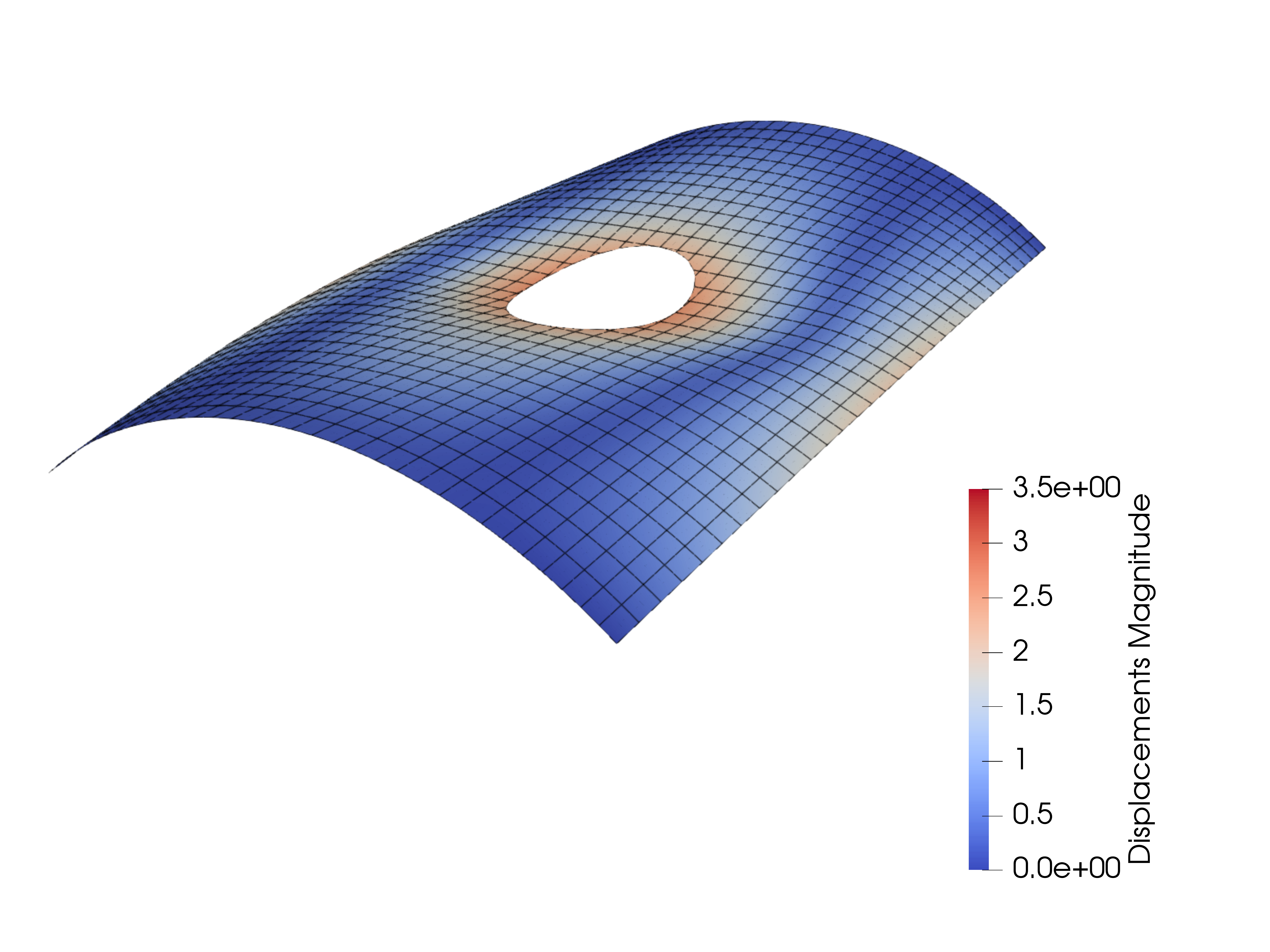}
		\caption{Load step 60.}
		\label{fig:scordelis_nonlin_step60}
	\end{subfigure}
	\begin{subfigure}{0.49\textwidth}
		\centering	
		\includegraphics[width=\textwidth,keepaspectratio]{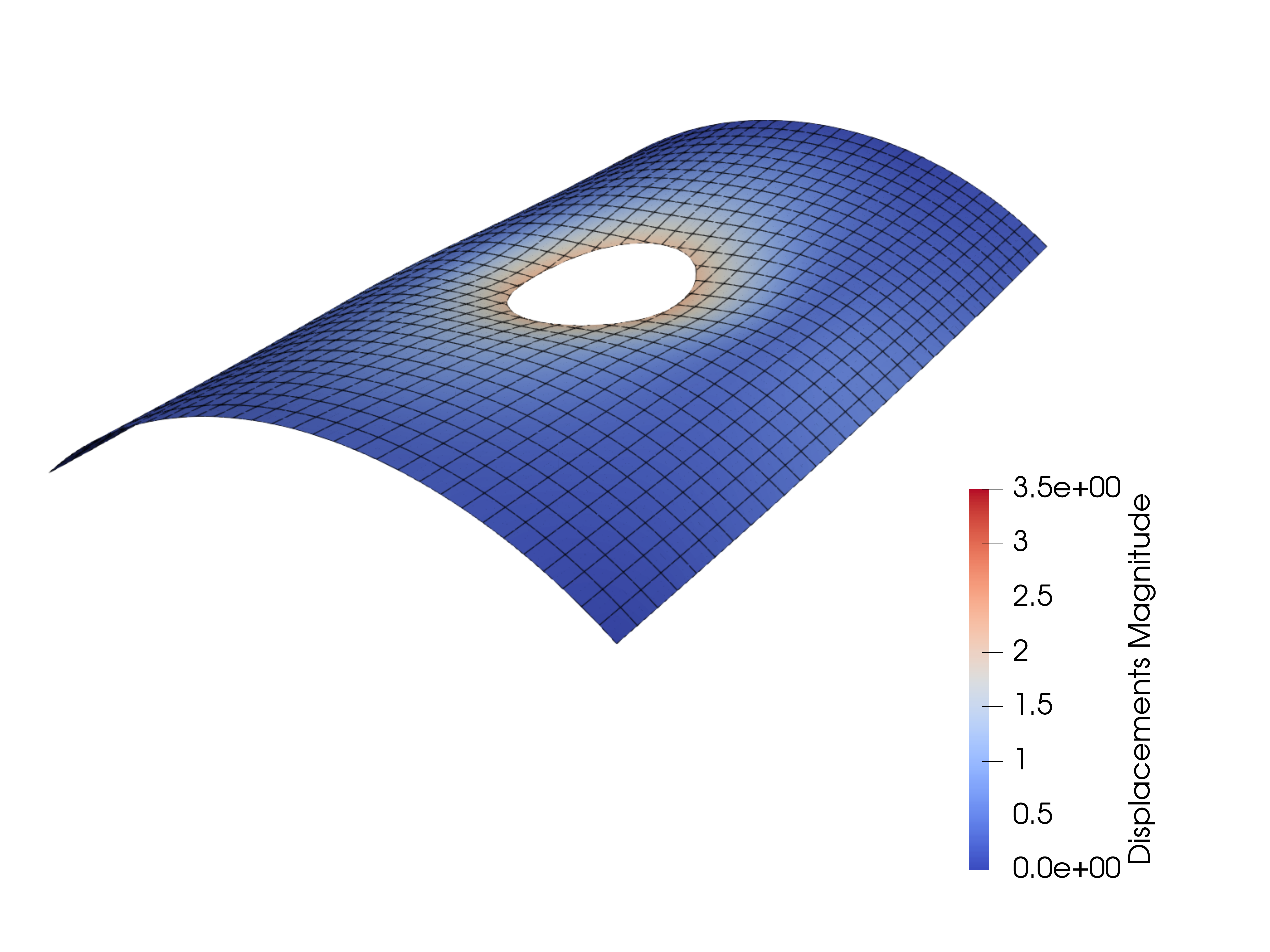}
		\caption{Load step 69 (last load step).}
		\label{fig:scordelis_nonlin_step69}
	\end{subfigure}
	\caption{Deformed Scordelis-Lo roof with elliptic hole illustrated in different load steps computed with $\sqrt{n_{ele}}=32$ and $p=3$. These results can be identically obtained by standard Gaussian and patch-wise quadrature.}
	\label{fig:scordelis_nonlin}
\end{figure}

The model is the same as in the previous Section \ref{sec:scordelis}, but a geometrically non-linear analysis is now performed. The load is increased to $q_z=900$. Fig. \ref{fig:scordelis_nonlin_load_displacement} shows the load-displacement curve. NURBS basis functions with a degree of $p=q=3$ are used. The proposed method is again compared with a standard trimming procedure. The results obtained by both methods perfectly match. The efficiency of this non-linear analysis profits from the reduction of the quadrature points as shown in Fig. \ref{fig:plate_reduction_GP} - in particular due to the fine meshes required for complex buckling patterns. Fig. \ref{fig:scordelis_nonlin} shows the deformation plots at different load steps and underlines the highly non-linear behaviour of the model.

\subsection{Geometrically Linear Intersecting Tubes}\label{sec:intersecting_tubes}
\begin{figure} [!bth]
	\centering
	\includegraphics[width=0.75\textwidth,keepaspectratio]{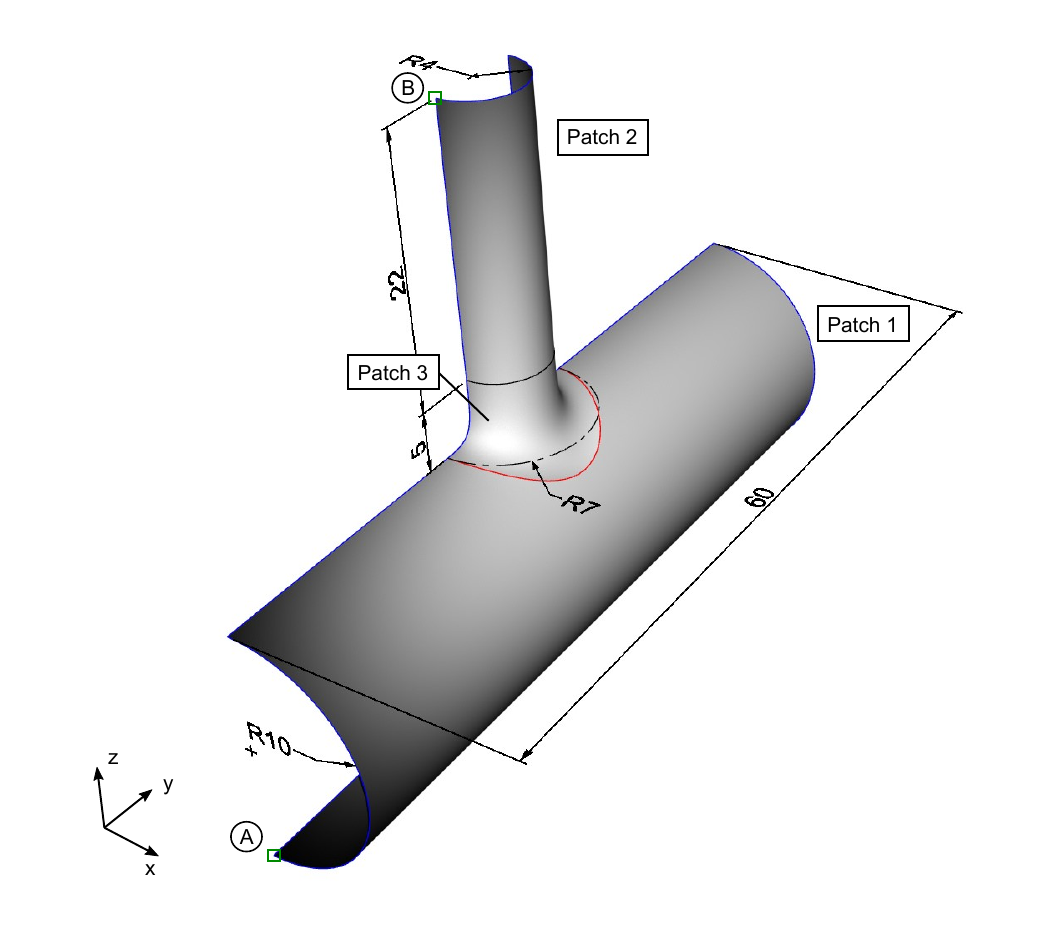}
	\caption{Setup of intersecting tubes. The black and the red curve are trimming curves and also indicate the coupling lines between the patches. Symmetry conditions are applied along the blue curves. The dotdashed line is a construction line used for the trimming of patch 1. The points A and B are used for the evaluation of results.}
	\label{fig:intersecting_tubes_system}
\end{figure}

This example demonstrates the abilities and advantages of the proposed method in the case of complex CAD geometries. The intersection of two cylindrical tubes connected by a fillet is investigated. The model was recently presented in \cite{Proserpio2022} and is created in the CAD program \textit{Rhinoceros 7} \cite{Mcneel}. For the simulation, the geometry, which was saved in the step format, is imported in \textit{Matlab} \cite{Matlab} by the open-source toolbox \cite{Loibl2023}. Fig. \ref{fig:intersecting_tubes_system} illustrates the geometry of the tubes. 

Patch 1, which is the larger cylinder (radius 10), is trimmed by a cylinder of radius 7 as indicated by the dot-dashed construction line. The resulting trimming curve is illustrated in red colour. Patch 2, which is the smaller cylinder (radius 4), is constructed by trimming a cylinder of height 37 with a x-y-plane in the height of the solid black line (resulting height is 22). Obviously, it could have been constructed without trimming since the resulting geometry is a cylinder again. Indeed, this is a good example where the principles of analysis-aware-modelling were not considered by the designer. However, it underlines the use of making patch-wise integration available for trimmed surfaces because trimming may be used even when it is not necessarily required when creating a certain geometry. The tubes are connected by a fillet, which is the untrimmed patch 3. Further geometrical details can be found in \cite{Proserpio2022}. The polynomial degree used for all patches is $p=q=4$. Patch 1 is refined with $40 \times 62$, patch 2 with $32 \times 40$ and patch 3 with $28 \times 19$ elements. Patch 2, which is trimmed as described above, actually contains only $32 \times 32$ active elements. Patch 1 and 3 are coupled along the red trimming curve, and patch 2 and 3 are coupled along the black trimming curve. Symmetry conditions are applied to all free edges, whereby the normals of the symmetry planes are the tangents of the shell at the respective edges. The coupling and symmetry conditions are applied by a penalty approach as presented in \cite{Proserpio2022} and \cite{Herrema2019}. The shell thickness is $t=0.2$. A linear elastic material with Young's modulus $E=3\cdot10^6$ and Poisson's ratio $\nu=0.3$ is considered. \bigskip

\begin{figure} [!bth]
	\centering
	\includegraphics[width=\textwidth,keepaspectratio]{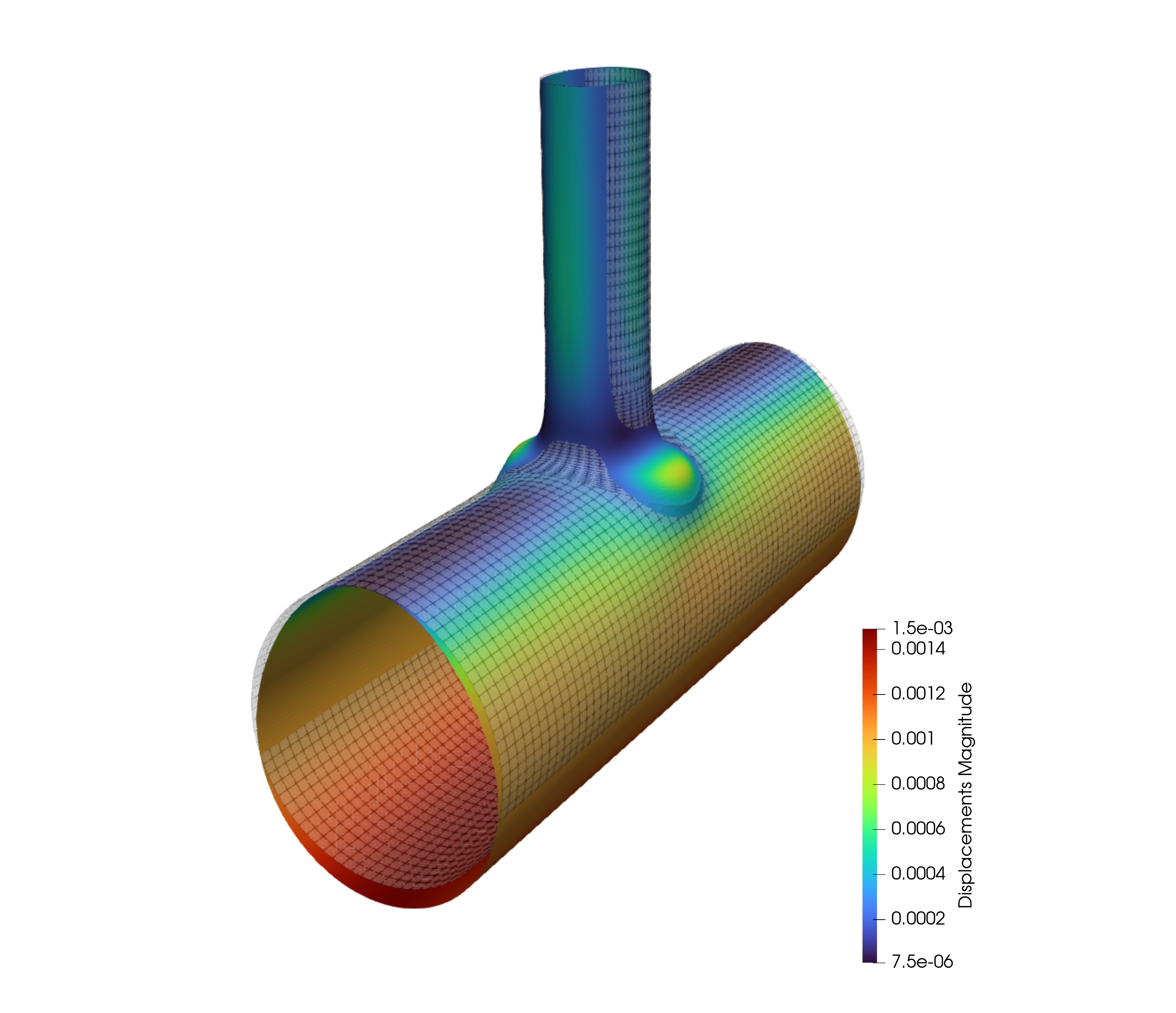}
	\caption{Deformed intersecting tubes under internal pressure on all patches. The undeformed geometry is illustrated as transparent grey mesh grid for reference. The depicted deformation is scaled by 1000.}
	\label{fig:intersecting_tubes_deformed}
\end{figure}

All patches are loaded by an internal pressure of 1. Fig. \ref{fig:intersecting_tubes_deformed} illustrates the deformation of the tubes. The displacement in $z$-direction at point A, which is depicted in Fig. \ref{fig:intersecting_tubes_system}, is equal to $u_{z,A} = 1.484\cdot10^{-3}$. The result is almost identical for the standard trimming and the new patch-wise approach with a relative difference of $2.12\cdot10^{-7}$. This result is in good agreement with the reference displacement $u_{z,A}=1.477\cdot10^{-3}$ by \cite{Proserpio2022} which is also computed with an isogeometric Kirchhoff-Love shell. \bigskip

The number of quadrature points can be reduced by approximately $60\%$ in this example. This underlines the high potential of speed-up for real-world problems by optimized quadrature rules. Table \ref{tab:intersecting_tubes} shows the number of quadrature points separately for all patches. Patch 3 contributes as expected the highest reduction of points with $51\%$, because it is an untrimmed patch and, therefore, the patch-wise integration rule shows its full potential. However, the number of points is also significantly decreased in patch 3 with $58\%$. The number of points in patch 2 could actually be reduced further if analysis-aware-modelling would be considered as discussed above.

\begin{table} 
	\caption{\label{tab:intersecting_tubes}Number of quadrature points for the example of the intersecting tubes.}
	\begin{center}
		\begin{tabular}{ccccc}
			& Total structure & Patch 1 & Patch 2 & Patch 3 \\ \hline &&&&\\[-2ex]
			$n_{quad,gauss}$ & 92500 & 60000 & 19200 & 13300 \\
			$n_{quad,pw}$ & 54619 & 34734 & 13153 & 6732 \\
			$n_{quad,pw}/n_{quad,gauss}$ & 0.590 & 0.579 & 0.685 & 0.506
		\end{tabular}
	\end{center}
\end{table}

\subsection{Geometrically Non-Linear Intersecting Tubes}
\begin{figure} [!bth]
	\centering
	\includegraphics{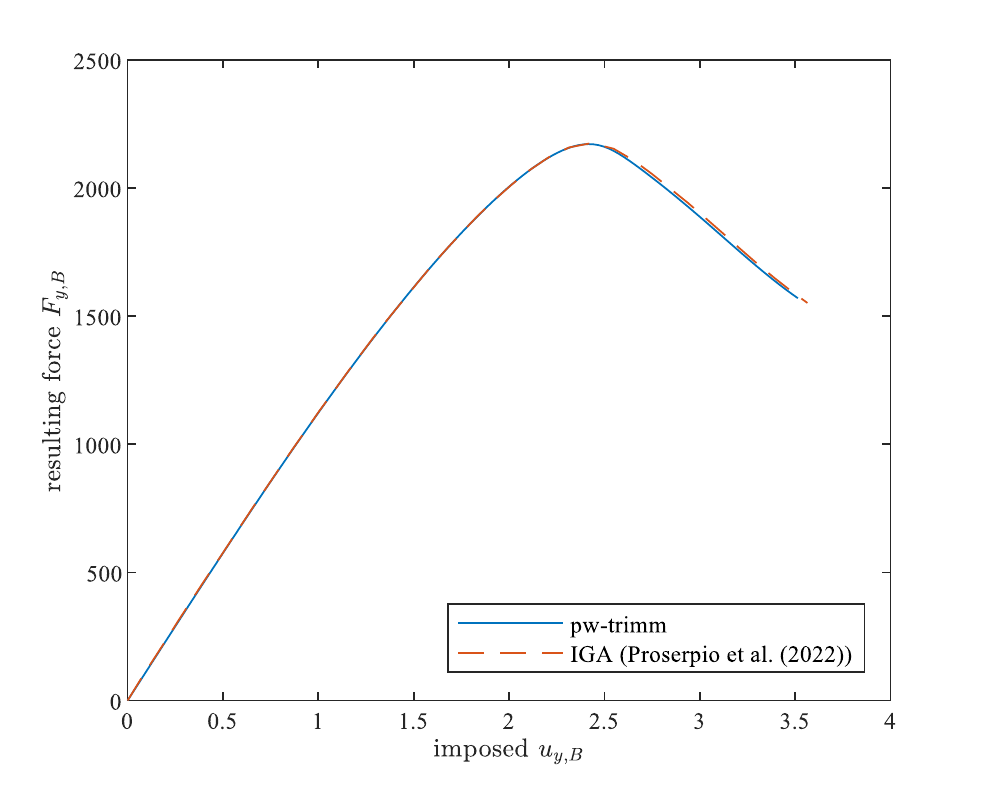}
	\caption{Load-displacement curve of the geometrically non-linear intersecting tubes. The result indicated with "Proserpio et al. (2022)" is extracted from \cite{Proserpio2022}.}
	\label{fig:intersecting_tubes_load_disp}
\end{figure}

In this Subsection, the same model as in the previous Section \ref{sec:intersecting_tubes} is considered but a geometrically non-linear analysis is now performed. The symmetry condition along the upper edge of patch 2 is now removed and instead a displacement in $y$-direction is imposed at this edge (see Fig. \ref{fig:intersecting_tubes_system}). The internal pressure applied in Section \ref{sec:intersecting_tubes} is removed. A point support in $z$-direction at point A is introduced in order to prohibit rigid body movements. Fig. \ref{fig:intersecting_tubes_load_disp} shows the resulting load-displacement curve for the force and displacement in $y$-direction at point B which is again in good agreement with the result from \cite{Proserpio2022} (see Fig. \ref{fig:intersecting_tubes_system} for the position of point B).

\section{Conclusion and Outlook}\label{sec:conclusion}
Patch-wise quadrature rules significantly reduce the number of quadrature points in the context of IGA. These methods are based on a tensor-product structure. Therefore, they are not directly applicable to trimmed structures which inherently violate the tensor-product. This publication presents the extension of patch-wise quadrature rules to trimmed surfaces. The key idea is that the elements are grouped in \textit{inactive}, \textit{trimmed}, \textit{transition} and \textit{patch-wise} elements, and that each group is integrated differently. In particular, the basis functions in the \textit{transition} elements are integrated in a mixed manner which enables a crossover from patch-wise to Gaussian quadrature points. \reviewed{The proposed method constructs the quadrature rule in a straightforward and unambiguous manner.} \bigskip

The method is successfully applied to plate in membrane and bending action, and shell problems. Geometrically linear and non-linear benchmarks are investigated. The possibility of a significant reduction of quadrature points is observed. In particular, a reduction of more than $50\%$ is noted in the example of the infinite plate with circular hole and a reduction of approximately $60\%$ is observed in the more practical example of the intersecting tubes. The results from the proposed method match well the results obtained from a standard trimming procedure underlying that the integration is performed equally good. \bigskip

Future work will concentrate on comparing studies of the different patch-wise quadrature rules. Special focus should be their performance with respect to integrands with rational terms. Existing research mostly investigates plane or solid problems with undistorted meshes, and neglects the quadrature errors arising from models with non-constant Jacobians due to distorted meshes or shell curvatures. \reviewed{Furthermore, an extension to local refinement techniques might be interesting because they are an efficient tool to locally resolve problems as fine as needed for a desired accuracy. Thereby, locally refined regions could be treated in a similar manner as trimmed parts. In addition, the varying number of quadrature points per element could lead to an unbalanced loading of the workers in parallelized code if an element-wise assembly is used. Possibilities in the assembly to alleviate this problem should be investigated such as direct iteration over quadrature points or splitting of mixed integrated elements which have the highest number of quadrature points. At last}, finding a set of optimized quadrature points for the \textit{transition} zone instead of a mixed integration would be a promising improvement for the proposed method \reviewed{and would lead to an optimal quadrature rule for trimmed structures}.

\section*{Declaration of competing interest}
The authors declare that they have no known competing financial interest or personal relationships that could have appeared to influence the work reported in this paper.

\section*{Acknowledgements}
\reviewed{This work} has received funding from the European Research Council (ERC) under the European Union's Horizon 2020 research and innovation program (grant agreement No 864482). Alessandro Reali was financially supported by the Italian Ministry for Education, University and Research (MIUR) through the PRIN project XFAST-SIMS (no. 20173C478N) and of the National Recovery and Resilience Plan, Mission 4 Component 2 -- Investment 1.4 -- CN\_00000013 CENTRO NAZIONALE ``HPC, BIG DATA E QUANTUM COMPUTING", spoke 6. All the support is gratefully acknowledged.

\appendix
\section{Displacement results for different examples}\label{app:displacement_results}
For the examples discussed in Sections \ref{sec:plate}, \ref{sec:punched_plate} and \ref{sec:scordelis}, displacement results are reported in Tab. \ref{tab:displacement_results}. The noted values are the converged values for the meshes depicted in the respective sections. Since Isogeometric Analysis is a displacement-based method, these results may facilitate faster reproducibility of the considered examples. For the examples from Sections \ref{sec:plate} and \ref{sec:scordelis}, the displacements are evaluated at a point A where this point is depicted in Fig. \ref{fig:plate_system} and \ref{fig:scordelis_elliptic_hole}, respectively. For the example from Section \ref{sec:punched_plate}, the displacement is evaluated at the centre of the plate.
\begin{table} 
	\caption{\label{tab:displacement_results}Displacement results for different examples}
	\begin{center}
		\begin{tabular}{ccc}
			{Example} & {Displacement} & {Value} \\ \hline &&\\[-2ex]
			Section \ref{sec:plate} & $u_{z,A}$ & -0.015113667 \\
			Section \ref{sec:punched_plate} & $u_{z,mid}$ & -0.002401 \\
			Section \ref{sec:scordelis} & $u_{z,A}$ & -0.3610789
		\end{tabular}
	\end{center}
\end{table}

\section{Geometric description of punched plate}\label{app:punched_plate_nurbs}
Here, the description of the NURBS trimming curves of the punched plate from Section \ref{sec:punched_plate} is provided. The trimming curves and the underlying rectangle patch are illustrated in Fig. \ref{fig:punched_plate_system}. The NURBS description of curve $C_1$ in the 2D physical space is:
\begin{align}
	CP_1 &= \left\{(3,2.5),(3,2),(2.5,2),(2,2),(2,2.5),(2,3),(2.5,3),(3,3),(3,2.5) \right\} \\
	w_1 &= \left\{1,\dfrac{1}{\sqrt{2}},1,\dfrac{1}{\sqrt{2}},1,\dfrac{1}{\sqrt{2}},1,\dfrac{1}{\sqrt{2}},1 \right\} \\
	\Xi_1 &= \left\{0,0,0,0.25,0.25,0.5,0.5,0.75,0.75,1,1,1 \right\} \\
	p_1 &= 2
\end{align}
The NURBS description of curve $C_2$ in the 2D physical space is:
\begin{align}
	CP_2 =& \{(2.5,8),(3.5,8),(4.5,8),(5,8),(5,7.5),(5,7),(4.5,7),(4,7),(3.5,7),(3,7),(3,6.5),(3,5.5), \nonumber \\
	& (3,5),(2.5,5),(2,5),(2,5.5),(2,6.5),(2,7.5),(2,8),(2.5,8)\} \\
	w_2 =& \left\{1,1,1,\dfrac{1}{\sqrt{2}},1,\dfrac{1}{\sqrt{2}},1,1,1,\dfrac{1}{\sqrt{2}},1,1,1,\dfrac{1}{\sqrt{2}},1,\dfrac{1}{\sqrt{2}},1,1,1,\dfrac{1}{\sqrt{2}},1 \right\} \\
	\Xi_2 =& \{0,0,0,0.1,0.1,0.2,0.2,0.3,0.3,0.4,0.4,0.5,0.5,0.6,0.6,0.7,0.7,0.8,0.8,0.9,0.9,1,1,1\} \\
	p =& 2
\end{align}
The NURBS description of curve $C_3$ in the 2D physical space is:
\begin{align}
	CP_3 =& \{(8,8),(9,8),(10,8),(10.5,8),(10.5,7.5),(10.5,7),(10,7),(9,7),(8,7),(7.5,7), \nonumber \\ & (7.5,7.5),(7.5,8),(8,8)\} \\
	w_2 =& \{1,1,1,\dfrac{1}{\sqrt{2}},1,\dfrac{1}{\sqrt{2}},1,1,1,\dfrac{1}{\sqrt{2}},1,\dfrac{1}{\sqrt{2}},1\} \\
	\Xi_3 =& \{0,0,0,0.25,0.25,0.375,0.375,0.5,0.5,0.75,0.75,0.875,0.875,1,1,1\} \\
	p =& 2
\end{align}

\bibliography{mybibfile}

\begin{thebibliography}{10}
\expandafter\ifx\csname url\endcsname\relax
  \def\url#1{\texttt{#1}}\fi
\expandafter\ifx\csname urlprefix\endcsname\relax\def\urlprefix{URL }\fi
\expandafter\ifx\csname href\endcsname\relax
  \def\href#1#2{#2} \def\path#1{#1}\fi

\bibitem{Hughes2005}
T.~J. Hughes, J.~A. Cottrell, Y.~Bazilevs, {Isogeometric analysis: CAD, finite
  elements, NURBS, exact geometry and mesh refinement}, Computer Methods in
  Applied Mechanics and Engineering 194~(39-41) (2005) 4135--4195.
\newblock \href {https://doi.org/10.1016/j.cma.2004.10.008}
  {\path{doi:10.1016/j.cma.2004.10.008}}.

\bibitem{Pan2020}
M.~Pan, B.~J{\"{u}}ttler, A.~Giust, {Fast formation of isogeometric Galerkin
  matrices via integration by interpolation and look-up}, Tech. Rep.~88, NFN -
  Nationales Forschungsnetzwerk (2020).
\newblock \href {https://doi.org/10.1016/j.cma.2020.113005}
  {\path{doi:10.1016/j.cma.2020.113005}}.

\bibitem{Auricchio2010}
F.~Auricchio, L.~B. {Da Veiga}, T.~J. Hughes, A.~Reali, G.~Sangalli,
  {Isogeometric collocation methods}, Mathematical Models and Methods in
  Applied Sciences 20~(11) (2010) 2075--2107.
\newblock \href {https://doi.org/10.1142/S0218202510004878}
  {\path{doi:10.1142/S0218202510004878}}.

\bibitem{Antolin2015}
P.~Antolin, A.~Buffa, F.~Calabr{\`{o}}, M.~Martinelli, G.~Sangalli, {Efficient
  matrix computation for tensor-product isogeometric analysis: The use of sum
  factorization}, Computer Methods in Applied Mechanics and Engineering 285
  (2015) 817--828.
\newblock \href {https://doi.org/10.1016/j.cma.2014.12.013}
  {\path{doi:10.1016/j.cma.2014.12.013}}.

\bibitem{Teschemacher2018}
T.~Teschemacher, A.~M. Bauer, T.~Oberbichler, M.~Breitenberger, R.~Rossi,
  R.~W{\"{u}}chner, K.-U. Bletzinger, {Realization of CAD-integrated shell
  simulation based on isogeometric B-Rep analysis}, Advanced Modeling and
  Simulation in Engineering Sciences 5~(19) (2018).
\newblock \href {https://doi.org/10.1186/s40323-018-0109-4}
  {\path{doi:10.1186/s40323-018-0109-4}}.

\bibitem{Leonetti2018}
L.~Leonetti, F.~Liguori, D.~Magisano, G.~Garcea, An efficient isogeometric
  solid-shell formulation for geometrically nonlinear analysis of elastic
  shells, Computer Methods in Applied Mechanics and Engineering 331 (2018) 159
  -- 183.
\newblock \href {https://doi.org/https://doi.org/10.1016/j.cma.2017.11.025}
  {\path{doi:https://doi.org/10.1016/j.cma.2017.11.025}}.

\bibitem{Calabro2019}
F.~Calabr{\`{o}}, G.~Loli, G.~Sangalli, M.~Tani, {Quadrature Rules in the
  Isogeometric Galerkin Method: State of the Art and an Introduction to
  Weighted Quadrature}, Springer INdAM Series 35 (2019) 43--55.
\newblock \href {https://doi.org/10.1007/978-3-030-27331-6_3}
  {\path{doi:10.1007/978-3-030-27331-6_3}}.

\bibitem{Hughes2010}
T.~J. Hughes, A.~Reali, G.~Sangalli, {Efficient quadrature for NURBS-based
  isogeometric analysis}, Computer Methods in Applied Mechanics and Engineering
  199~(5-8) (2010) 301--313.
\newblock \href {https://doi.org/10.1016/j.cma.2008.12.004}
  {\path{doi:10.1016/j.cma.2008.12.004}}.

\bibitem{Auricchio2012}
F.~Auricchio, F.~Calabr{\`{o}}, T.~J. Hughes, A.~Reali, G.~Sangalli,
  \href{http://dx.doi.org/10.1016/j.cma.2012.04.014}{{A simple algorithm for
  obtaining nearly optimal quadrature rules for NURBS-based isogeometric
  analysis}}, Computer Methods in Applied Mechanics and Engineering 249-252
  (2012) 15--27.
\newblock \href {https://doi.org/10.1016/j.cma.2012.04.014}
  {\path{doi:10.1016/j.cma.2012.04.014}}.
\newline\urlprefix\url{http://dx.doi.org/10.1016/j.cma.2012.04.014}

\bibitem{Adam2015}
C.~Adam, T.~J. Hughes, S.~Bouabdallah, M.~Zarroug, H.~Maitournam, {Selective
  and reduced numerical integrations for NURBS-based isogeometric analysis},
  Computer Methods in Applied Mechanics and Engineering 284 (2015) 732--761.
\newblock \href {https://doi.org/10.1016/j.cma.2014.11.001}
  {\path{doi:10.1016/j.cma.2014.11.001}}.

\bibitem{Johannessen2017}
K.~A. Johannessen, {Optimal quadrature for univariate and tensor product
  splines}, Computer Methods in Applied Mechanics and Engineering 316 (2017)
  84--99.
\newblock \href {https://doi.org/10.1016/j.cma.2016.04.030}
  {\path{doi:10.1016/j.cma.2016.04.030}}.

\bibitem{Hiemstra2017}
R.~R. Hiemstra, F.~Calabr{\`{o}}, D.~Schillinger, T.~J. Hughes, {Optimal and
  reduced quadrature rules for tensor product and hierarchically refined
  splines in isogeometric analysis}, Computer Methods in Applied Mechanics and
  Engineering 316 (2017) 966--1004.
\newblock \href {https://doi.org/10.1016/j.cma.2016.10.049}
  {\path{doi:10.1016/j.cma.2016.10.049}}.

\bibitem{Calabro2017}
F.~Calabr{\`{o}}, G.~Sangalli, M.~Tani, {Fast formation of isogeometric
  Galerkin matrices by weighted quadrature}, Computer Methods in Applied
  Mechanics and Engineering 316 (2017) 606--622.
\newblock \href {http://arxiv.org/abs/arXiv:1605.01238v2}
  {\path{arXiv:arXiv:1605.01238v2}}, \href
  {https://doi.org/10.1016/j.cma.2016.09.013}
  {\path{doi:10.1016/j.cma.2016.09.013}}.

\bibitem{Sangalli2018}
G.~Sangalli, M.~Tani, {Matrix-free weighted quadrature for a computationally
  efficient isogeometric k-method}, Computer Methods in Applied Mechanics and
  Engineering 338 (2018) 117--133.
\newblock \href {http://arxiv.org/abs/1712.08565} {\path{arXiv:1712.08565}},
  \href {https://doi.org/10.1016/j.cma.2018.04.029}
  {\path{doi:10.1016/j.cma.2018.04.029}}.

\bibitem{Hiemstra2019}
R.~Hiemstra, G.~Sangalli, M.~Tani, F.~Calabr{\`{o}}, T.~J.~R. Hughes, {Fast
  Formation and Assembly of Finite Element Matrices with Application to
  Isogeometric Linear Elasticity}, Computer Methods in Applied Mechanics and
  Engineering 355 (2019) 234--260.
\newblock \href {https://doi.org/10.1016/j.cma.2019.06.020}
  {\path{doi:10.1016/j.cma.2019.06.020}}.

\bibitem{Giannelli2021}
C.~Giannelli, T.~Kanduc, M.~Martinelli, G.~Sangalli, M.~Tani, {Weighted
  quadrature for hierarchical B-splines}, Computer Methods in Applied Mechanics
  and Engineering 400 (2022).
\newblock \href {http://arxiv.org/abs/2109.12632} {\path{arXiv:2109.12632}},
  \href {https://doi.org/10.1016/j.cma.2022.115465}
  {\path{doi:10.1016/j.cma.2022.115465}}.

\bibitem{Leonetti2019}
L.~Leonetti, D.~Magisano, A.~Madeo, G.~Garcea, J.~Kiendl, A.~Reali, {A
  simplified Kirchhoff–Love large deformation model for elastic shells and
  its effective isogeometric formulation}, Computer Methods in Applied
  Mechanics and Engineering 354 (2019) 369--396.
\newblock \href {https://doi.org/10.1016/j.cma.2019.05.025}
  {\path{doi:10.1016/j.cma.2019.05.025}}.

\bibitem{Hokkanen2020}
J.~Hokkanen, D.~M. Pedroso, {Quadrature rules for isogeometric shell
  formulations: Study using a real-world application about metal forming},
  Computer Methods in Applied Mechanics and Engineering 363~(112904) (2020).
\newblock \href {https://doi.org/10.1016/j.cma.2020.112904}
  {\path{doi:10.1016/j.cma.2020.112904}}.

\bibitem{Hokkanen2019}
J.~Hokkanen, {Isogeometric shell analysis of incremental sheet forming}, Ph.D.
  thesis, University of Queensland (2019).
\newblock \href {https://doi.org/10.14264/uql.2020.116}
  {\path{doi:10.14264/uql.2020.116}}.

\bibitem{Breitenberger2016}
M.~Breitenberger, {CAD-Integrated Design and Analysis of Shell Structures},
  Ph.D. thesis, Technical University Munich (2016).

\bibitem{Breitenberger2015}
M.~Breitenberger, A.~Apostolatos, B.~Philipp, R.~W{\"{u}}chner, K.-U.
  Bletzinger, {Analysis in computer aided design: Nonlinear isogeometric B-Rep
  analysis of shell structures}, Computer Methods in Applied Mechanics and
  Engineering 284 (2015) 401--457.
\newblock \href {https://doi.org/10.1016/j.cma.2014.09.033}
  {\path{doi:10.1016/j.cma.2014.09.033}}.

\bibitem{Marussig2018}
B.~Marussig, T.~J. Hughes, {A Review of Trimming in Isogeometric Analysis:
  Challenges, Data Exchange and Simulation Aspects}, Archives of Computational
  Methods in Engineering 25~(4) (2018) 1059--1127.
\newblock \href {https://doi.org/10.1007/s11831-017-9220-9}
  {\path{doi:10.1007/s11831-017-9220-9}}.

\bibitem{Schmidt2012}
R.~Schmidt, R.~W{\"{u}}chner, K.~U. Bletzinger, {Isogeometric analysis of
  trimmed NURBS geometries}, Computer Methods in Applied Mechanics and
  Engineering 241-244 (2012) 93--111.
\newblock \href {https://doi.org/10.1016/j.cma.2012.05.021}
  {\path{doi:10.1016/j.cma.2012.05.021}}.

\bibitem{Nagy2015}
A.~P. Nagy, D.~J. Benson, {On the numerical integration of trimmed isogeometric
  elements}, Computer Methods in Applied Mechanics and Engineering 284 (2015)
  165--185.
\newblock \href {https://doi.org/10.1016/j.cma.2014.08.002}
  {\path{doi:10.1016/j.cma.2014.08.002}}.

\bibitem{Messmer2022}
M.~Me{\ss}mer, T.~Teschemacher, L.~Leidinger, R.~W{\"{u}}chner, K.-U.
  Bletzinger, {Efficient CAD-integrated isogeometric analysis of trimmed
  solids}, Computer Methods in Applied Mechanics and Engineering 400~(115584)
  (2022) 1--47.
\newblock \href {https://doi.org/10.1016/j.cma.2022.115584}
  {\path{doi:10.1016/j.cma.2022.115584}}.

\bibitem{Rank2012}
E.~Rank, M.~Ruess, S.~Kollmannsberger, D.~Schillinger, A.~D{\"{u}}ster,
  {Geometric modeling, isogeometric analysis and the finite cell method},
  Computer Methods in Applied Mechanics and Engineering 249-252 (2012)
  104--115.
\newblock \href {https://doi.org/10.1016/j.cma.2012.05.022}
  {\path{doi:10.1016/j.cma.2012.05.022}}.

\bibitem{Kim2009}
H.~J. Kim, Y.~D. Seo, S.~K. Youn, {Isogeometric analysis for trimmed CAD
  surfaces}, Computer Methods in Applied Mechanics and Engineering 198~(37-40)
  (2009) 2982--2995.
\newblock \href {https://doi.org/10.1016/j.cma.2009.05.004}
  {\path{doi:10.1016/j.cma.2009.05.004}}.

\bibitem{Kim2010}
H.~J. Kim, Y.~D. Seo, S.~K. Youn, {Isogeometric analysis with trimming
  technique for problems of arbitrary complex topology}, Computer Methods in
  Applied Mechanics and Engineering 199~(45-48) (2010) 2796--2812.
\newblock \href {https://doi.org/10.1016/j.cma.2010.04.015}
  {\path{doi:10.1016/j.cma.2010.04.015}}.

\bibitem{Guo2018}
Y.~Guo, J.~Heller, T.~J. Hughes, M.~Ruess, D.~Schillinger, {Variationally
  consistent isogeometric analysis of trimmed thin shells at finite
  deformations, based on the STEP exchange format}, Computer Methods in Applied
  Mechanics and Engineering 336 (2018) 39--79.
\newblock \href {https://doi.org/10.1016/j.cma.2018.02.027}
  {\path{doi:10.1016/j.cma.2018.02.027}}.

\bibitem{Leonetti2018a}
L.~Leonetti, D.~Magisano, F.~Liguori, G.~Garcea, An isogeometric formulation of
  the koiter’s theory for buckling and initial post-buckling analysis of
  composite shells, Computer Methods in Applied Mechanics and Engineering 337
  (2018) 387 -- 410.
\newblock \href {https://doi.org/https://doi.org/10.1016/j.cma.2018.03.037}
  {\path{doi:https://doi.org/10.1016/j.cma.2018.03.037}}.

\bibitem{Marussig2022}
B.~Marussig, {Fast formation and assembly of isogeometric Galerkin matrices for
  trimmed patches}, in: C.~Manni, H.~Speleers (Eds.), Geometric challenges in
  Isogeometric Analysis, Springer Nature Switzerland AG, Cham, 2022, pp.
  149--170.
\newblock \href {https://doi.org/10.1007/978-3-030-92313-6_7}
  {\path{doi:10.1007/978-3-030-92313-6_7}}.

\bibitem{Piegl1995}
L.~Piegl, W.~Tiller, {The NURBS Book}, Springer Berlin Heidelberg, Berlin,
  Heidelberg, 1995.

\bibitem{Kiendl2011}
J.~M. Kiendl, {Isogeometric Analysis and Shape Optimal Design of Shell
  Structures}, Ph.D. thesis, Technical University Munich (2011).

\bibitem{Dornisch2021}
W.~Dornisch, J.~St{\"{o}}ckler, {An isogeometric mortar method for the coupling
  of multiple NURBS domains with optimal convergence rates}, Numerische
  Mathematik 149~(4) (2021) 871--931.
\newblock \href {https://doi.org/10.1007/s00211-021-01246-z}
  {\path{doi:10.1007/s00211-021-01246-z}}.

\bibitem{Szabo1991}
B.~Szab{\'{o}}, I.~Babu{\v{s}}ka, {Finite Element Analysis}, John Wiley {\&}
  Sons, Inc., New York, Chichester, Brisbane, Toronto, Singapur, 1991.

\bibitem{Antolin2019}
P.~Antolin, A.~Buffa, M.~Martinelli, {Isogeometric Analysis on V-reps: First
  results}, Computer Methods in Applied Mechanics and Engineering 355 (2019)
  976--1002.
\newblock \href {http://arxiv.org/abs/1903.03362} {\path{arXiv:1903.03362}},
  \href {https://doi.org/10.1016/j.cma.2019.07.015}
  {\path{doi:10.1016/j.cma.2019.07.015}}.

\bibitem{Coradello2020b}
L.~Coradello, P.~Antolin, R.~V{\'{a}}zquez, A.~Buffa, {Adaptive isogeometric
  analysis on two-dimensional trimmed domains based on a hierarchical
  approach}, Computer Methods in Applied Mechanics and Engineering 364 (2020)
  1--25.
\newblock \href {http://arxiv.org/abs/1910.03882} {\path{arXiv:1910.03882}},
  \href {https://doi.org/10.1016/j.cma.2020.112925}
  {\path{doi:10.1016/j.cma.2020.112925}}.

\bibitem{Coradello2020a}
L.~Coradello, D.~D'Angella, M.~Carraturo, J.~Kiendl, S.~Kollmannsberger,
  E.~Rank, A.~Reali,
  \href{https://doi.org/10.1007/s00466-020-01858-6}{{Hierarchically refined
  isogeometric analysis of trimmed shells}}, Computational Mechanics 66~(2)
  (2020) 431--447.
\newblock \href {https://doi.org/10.1007/s00466-020-01858-6}
  {\path{doi:10.1007/s00466-020-01858-6}}.
\newline\urlprefix\url{https://doi.org/10.1007/s00466-020-01858-6}

\bibitem{Proserpio2022}
D.~Proserpio, J.~Kiendl, {Penalty coupling of trimmed isogeometric
  Kirchhoff-Love shell patches}, Journal of Mechanics 38~(February) (2022)
  156--165.
\newblock \href {https://doi.org/10.1093/jom/ufac008}
  {\path{doi:10.1093/jom/ufac008}}.

\bibitem{Mcneel}
M.~. Associates, \href{https://www.rhino3d.com/7/}{Rhinoceros 7} (2022).
\newline\urlprefix\url{https://www.rhino3d.com/7/}

\bibitem{Matlab}
I.~The~MathWorks, \href{https://de.mathworks.com/products/matlab.html}{Matlab
  2019b} (2019).
\newline\urlprefix\url{https://de.mathworks.com/products/matlab.html}

\bibitem{Loibl2023}
M.~Loibl,
  \href{https://www.mathworks.com/matlabcentral/fileexchange/123425-step-file-import-of-trimmed-nurbs-surfaces}{Step
  file import of trimmed nurbs surfaces} (2023).
\newline\urlprefix\url{https://www.mathworks.com/matlabcentral/fileexchange/123425-step-file-import-of-trimmed-nurbs-surfaces}

\bibitem{Herrema2019}
A.~J. Herrema, E.~L. Johnson, D.~Proserpio, M.~C. Wu, J.~Kiendl, M.~C. Hsu,
  {Penalty coupling of non-matching isogeometric Kirchhoff–Love shell patches
  with application to composite wind turbine blades}, Computer Methods in
  Applied Mechanics and Engineering 346 (2019) 810--840.
\newblock \href {https://doi.org/10.1016/j.cma.2018.08.038}
  {\path{doi:10.1016/j.cma.2018.08.038}}.

\end{thebibliography}

\end{document}